\documentclass{article}

\usepackage{amssymb}
\usepackage{amsthm}
\usepackage{amsmath}

\usepackage{graphicx}

\newtheorem{theorem}{Theorem}[section]
\newtheorem{prop}[theorem]{Proposition}
\newtheorem{lem}[theorem]{Lemma}

\newtheorem{remark}[theorem]{Remark}

\newcommand{\ou}{{\overline{u}}}
\newcommand{\ov}{{\overline{v}}}
\newcommand{\ox}{{\overline{x}}}
\newcommand{\ot}{{\overline{t}}}
\newcommand{\D}{{\mathrm{d}}}

\begin{document}
\title{On the plane into plane mappings of hydrodynamic type. Parabolic case.}
\author{B.G. Konopelchenko \\
Dipartimento di Matematica e Fisica ``Ennio de Giorgi", Universit\`{a} del
Salento \\
and INFN, Sezione di Lecce, 73100 Lecce, Italy \\
{konopel@le.infn.it} \\
\mbox{} \\
G. Ortenzi \\
Dipartimento di Matematica Pura ed Applicazioni,\\
Universit\`{a} di Milano Bicocca, 20125 Milano, Italy\\
{giovanni.ortenzi@unimib.it} }
\maketitle
\abstract{Singularities of plane into plane mappings described by parabolic two-component systems of quasi-liner partial differential equations of the first order are
studied. Impediments arising in the application of the original Whitney's approach to such case are discussed. Hierarchy of singularities is analysed by double-scaling expansion method for the simplest $2$-component Jordan system. It is shown that flex is the lowest singularity while higher singularities are given  by 
$(k+1,k+2)$ curves which are of cusp type for  $k=2n+1$, $n=1,2,3,\dots$. Regularization of these singularities by deformation of plane into plane mappings into surface $S^{2+k}  (\subset \mathbb{R}^{2+k} ) $  to plane is discussed. Applicability of the proposed approach to other parabolic type mappings is noted. }
\vspace{1cm}
\begin{flushright}
{\it To the memory of B. A. Dubrovin}
\end{flushright}

\section{Introduction}
Singularities of solutions of hyperbolic and elliptic partial differential equations (PDEs)
and associated mappings have been intensively studied during last decades
(see e.g. \cite{CF,Bers,Hor,GS,Guc,Kri,Lyc1,Lyc2,Rak1,Rak2,Giv,KS,Sul,KMAM,Dub1,Dub2,KOSapm,DGKM} and references therein). 
Parabolic case, viewed as the degenerate
case, has attracted much less attention. It has been addressed essentially only
within the study of behavior of solutions of PDEs of mixed type near the transition
(sonic) line in fluid- and gas-dynamics (see e.g. \cite{Bers,L-VI}) and for heat and diffusion type PDEs of second order \cite{Sul,Ale,Mag}.\\

  Mappings related to the systems of quasi-linear PDEs of the first order  represent
  particular class of plane into plane mappings. However due to the existence of hodograph transformations (see e.g. \cite{CF,Bers,L-VI}) such equations provide us, 
  probably, with the best laboratory for testing the singularity properties of associated mappings and corresponding solutions. Until now such an analysis
  has been successfully performed for hyperbolic systems (see \cite{Hor,GS,Guc,Kri,Lyc1,Lyc2,Rak1,Rak2,Giv,KS,Sul,KMAM,Dub1,Dub2}).  \par
  
  Peculiarity of parabolic regime has been noted in B. Dubrovin fundamental paper on critical behaviour in Hamiltonian PDEs \cite{Dub2}, but was not really addressed.
  Our aim is to fill partially this gap. 
  
   In the present paper we will study plane into plane mappings and their singularities governed 
  by the parabolic system
  \begin{equation}
 \binom{u_t}{v_t}=\left( \begin{array}{cc}\lambda & 1 \\ 0 & \lambda \end{array} \right) \binom{u_x}{v_x}
 \label{genparasys}
\end{equation}
  where $\lambda$ is a function of $u$ and $v$. 
System (\ref{genparasys}) belongs to the class of integrable  quasilinear systems of hydrodynamic type \cite{KK}. 
Systems of the type (\ref{genparasys})
arise also on the transition line for several systems of PDEs of mixed type, for instance, those describing plane motion in gas and fluid dynamics (see e.g. \cite{KOJor,CFO}). \par

Standard hodograph equations for the system (\ref{genparasys}), i.e. the system
\begin{equation}
x_u+\lambda t_u=0\, , \qquad x_v+\lambda t_v=t_u\, , 
\label{hodogenparasys}
\end{equation} 
provides us with the $\mathbb{R}^2 \to \mathbb{R}^2$ mappings $(u,v)\to (t,x)$. 
In the particular case $\lambda=u$ such mappings have simple explicit form
\begin{equation}
t=W_u\, , \qquad x= W_u-u W_v\, ,
\label{solusys}
\end{equation}
where  function $W_{u,v}$ is a solution of the heat equation
\begin{equation}
W_v=W_{vv}\, .
\label{heateqn}
\end{equation}

This class of $\mathbb{R}^2 \to \mathbb{R}^2$ mappings will be studied in detail 
in the present paper. 
First, it shown that particular features of the parabolic mappings (\ref{solusys})-(\ref{heateqn}) and those associated with 
the system (\ref{genparasys}) prevent the direct and effective application of the original 
Whitney approach \cite{W} to them.  Some important properties of generic mappings \cite{W} are no more 
valid for such parabolic mappings. In particular,  
there are no ``excellent'' mappings, i.e. those containing only folds and standard $(2,3)$ cusps \cite{W}. \par

 Second, we analyze not only the first order singularities of the mappings  
 (\ref{solusys})-(\ref{heateqn}), but the whole infinite family of singularities. In contrast to the usually adopted approach to consider the ``generic'' case
 (see e.g. \cite{Hor,GS,Guc,Kri,Lyc1,Lyc2,Rak1,Rak2,Giv,KS,Sul,Dub1,Dub2}) we follow the general principle formulated by Poincar\'e \cite{Poi}
 according to which  ``one has to study not only a single situation (even generic one) but the whole family of close situations in order to get complete and deep understanding of certain phenomenon''.  In our case it means  that we have to consider a  family of mappings (\ref{solusys}) for an
 infinite family of solutions $W$ of the equation (\ref{heateqn}) which corresponds to family of solutions of
 the system (\ref{hodogenparasys}). Higher singularities $(\ref{solusys})$  correspond to the degenerate critical points of higher order for the function $W$ obeying (\ref{heateqn}). So, higher order singularities are unremovable for the family of mappings (\ref{solusys}) similar to the degenerate critical points of higher order which are not
 removable for families of functions (see e.g. \cite{Thom,Arn1,Gol}).\par
 Using the double scaling expansion technique we show that near the singularity of the order $k$ $(k=1,2,3,\dots)$ the mapping (\ref{solusys}) has locally the form
 \begin{equation}
 \overline{t}_k = P_{k+1}(\overline{u},\overline{v})\, ,\qquad 
 \overline{x}_k = P_{k+2}(\overline{u},\overline{v}) - \overline{u} P_{k+1}(\overline{u},\overline{v})
 \, ,
 \label{locgensing}
 \end{equation}
where $P_{k}(\overline{u},\overline{v})$ are elementary Schur polynomials of two variables. For $k=1$ the mapping (\ref{locgensing}) is the flex 
\begin{equation}
\overline{t_1} =\frac{1}{2} \overline{u}^2+\overline{v}\, , \qquad 
\overline{x_1} =\frac{1}{3} \overline{u}^3\, , 
\label{fold}
\end{equation}
while at $k=2$ it is
\begin{equation}
\overline{t}_2 =\frac{1}{6} \overline{u}^3+\overline{u}\, \overline{v}\, , \qquad 
\overline{x}_2 =-\frac{1}{8} \overline{u}^4 -\frac{1}{2} \overline{u}^2\overline{v}
+\frac{1}{2} \overline{v}^2\, .
\label{supercusp}
\end{equation}
The mappings (\ref{locgensing}) are singular along the curve $P_{k}(\overline{u},\overline{v})=0$. It is shown that this curve is the union of $n$ parabolas 
$\overline{u}^2 + 4 \alpha_i \overline{v}=0$, $i=1,\dots, n$  for $k=2n$ and the union of the line $\overline{u}=0$ and the $n$ parabolas 
$\overline{u}^2 + 4 \alpha_i \overline{v}=0$, $i=1,\dots, n$  for $k=2n+1$, where $\alpha_i$ are roots of Hermite polynomials.
Straight lines $\ox_{2n+1}=0$ are images of the line $\ou=0$. Images of parabolas are $(k+1,k+2)$ curves which have  at origin ($\ot=\ox=0$) $k$-th order singular points with unbounded curvature. For $k=2n+1$, $n=1,2,3,\dots$ these curves are  $(2n+2,2n+3)$ cusps. \par

We discuss also a regularization of singularities of mappings (\ref{solusys})-(\ref{heateqn}) by deforming them into the mapping of certain surfaces $S^N(u)$
in $\mathbb{R}^N$ into the plane $(t,x)$. It is shown that $k$-th order singularities of mappings (\ref{solusys})-(\ref{heateqn}) are regularized by deforming them  
into mappings $S^{k+2}(u) \to (t,x)$-plane. \par

Applicability of the approach presented in the paper to an analysis of singularities associated with other parabolic systems of PDEs is briefly discussed. 
Singularities of above $S^{2+k}(u)\to (t,x)$-plane mappings, $\mathbb{R}^2 \to \mathbb{R}^2$ mappings describing symmetries of the system 
(\ref{genparasys}) and certain    $\mathbb{R}^n \to \mathbb{R}^n$ mappings, associated with various integrable parabolic extensions of the system (\ref{genparasys})
are among them. \par

In order to emphasize the peculiarity of the parabolic mappings we consider briefly the mappings $(u,v)\to(t,x)$ for hyperbolic two-component systems of
hydrodynamic type. It is shown that the Whitney classification approach is an effective one in this case. Higher singularities are also analyzed. \par

The paper is organized as follows: Section \ref{Whitapp-sec}  is devoted to the brief recall of the original Whitney's approach. In section 
\ref{Whitob-sec} the impediments for application of Whitney's approach to the mappings governed by the system (\ref{genparasys}) are discussed. The mappings (\ref{solusys})-(\ref{heateqn}) and their singularities are studied in section \ref{Jorsing-sec}. Structure of singular curves  for mapping (\ref{locgensing})  is analyzed in section \ref{singcurv-sec}. Regularization of singularities for the mappings (\ref{locgensing}) is described in section \ref{regul-sec}.
 Other parabolic type mappings treatable in the same way are mentioned in section \ref{otherpara-sec}. 
Application of the  Whitney's approach to hyperbolic system is discussed in section \ref{hyper-sec}. 
\section{Whitney's theory of singularities}
\label{Whitapp-sec}
For convenience we briefly recall here some basic facts from the original Whitney's analysis of singularities of the plane into the plane mappings \cite{W}
(see also \cite{Arn1,Rem2}) with notations adapted to our considerations.\par

Let $f$ be a good $2$-smooth mapping $(u,v)\to (t,x)$, i.e. such that at every point of the plane $(u,v)$ either the Jacobian
\begin{equation}
J=t_u x_v-t_v x_u
\label{Jc2}
\end{equation}
is different from zero or at least one of $J_u$ and $J_v$ is different from zero \cite{W}. Singular points of $f$ belongs to the curve $J=0$. Due to the ``goodness''
condition this curve is smooth and at each point $p$ there is nonzero tangent vector $\vec{V} = (-J_v(p),J_u(p))$. Considering the differentiation along this vector, 
one introduces the vector field 
 \begin{equation}
 \nabla_V = -J_v \partial_u+J_u \partial_v \, ,
 \label{tanvec}
 \end{equation}
 which acts on the plane $(t,x)$ according to the formula  
 \begin{equation}
 \nabla_V f(p)= -J_v f_u(p)+J_u f_v (p)\, .
 \label{tanvec-app}
 \end{equation}
The point $p$ of the singular curve $J=0$ is a fold point if and only if \cite{W}
 \begin{equation}
 \nabla_V f(p)\neq 0 \, ,
 \label{foldcond}
 \end{equation}
and  $p$ is a cusp point if and only if \cite{W}
 \begin{equation}
 \nabla_V f(p)= 0 \, , \qquad \nabla_V^2 f(p)\neq 0\, .
 \label{cuspcond}
 \end{equation}
 It was shown \cite{W} that near the fold the mapping $f$ has a normal form
 \begin{equation}
 t=v\, , \qquad x=u^2
 \label{genfold}
 \end{equation}
 while near the cusp  it is of the form 
 \begin{equation}
 t=v\, , \qquad x=u^3-uv\, .
 \label{gencusp}
 \end{equation}
The mapping $f$ has been called ``excellent'' \cite{W} if all its singular points are fold or cusp points.
Then the basic theorem $13$A in \cite{W} (p.$389$) states that ``arbitrarily near any mapping $f_0$ there is an excellent mapping $f$''.
A crucial condition for this theorem to be valid is that the bad set $S$ defined by the three equations (formula (13.5)  \cite{W} p.$390$)
\begin{equation}
J=0\, ,\qquad \nabla_V f=0\, , \qquad  \nabla_{V}^{2} f=0\, ,
\label{3Wcond}
\end{equation}
  has the defect (codimension) $3$, i.e. the equations (\ref{3Wcond}) are independent.  In addition, Whitney has chosen the coordinates systems in the
  planes $(u,v)$ and $(t,x)$ characterized by the conditions 
   \begin{equation}
   t_u=t_v=x_u=0\, , \quad x_v=1 \qquad \mathrm{at\, } p\, .
   \label{Wnorm}
   \end{equation}
Such a choice was quite convenient for simplification fo calculations in \cite{W}, but, in fact, it is not essential one (see e.g. \cite{Rem2}).
\section{Generic parabolic case: impediments}
\label{Whitob-sec}
We start with the generic parabolic two-component hydrodynamic type system with two independent variables. It can be reduced to the form  \cite{KOJor}
\begin{equation}
u_t= \lambda u_x +v_x\, , \qquad v_t = \lambda u_x\, , 
\label{genparasys2}
\end{equation}
where $\lambda$ is a certain function of $u$ and $v$. We will assume that $\lambda_u \neq 0$, otherwise the system   (\ref{genparasys2}) is totally decoupled.\par

In the hodograph plane $(u,v)$ the system (\ref{genparasys2}) is of the form (see e.g. \cite{KOJor})
\begin{equation}
\label{Hodoeqn-2par}
 x_u = -\lambda t_u  \, \qquad x_v = t_u - \lambda t_v \, .  \end{equation}
This system is compatible iff 
\begin{equation}  
 \label{t-2parcomp}
  t_{uu}=\lambda_u t_v-\lambda_v t_u \, .
\end{equation}
Equations (\ref{Hodoeqn-2par}) are equivalent to
\begin{equation}
\label{Hodo2par}
  (x+\lambda\, t)_u =\lambda_u t  \, \qquad  (x+\lambda\, t)_v = t_u +\lambda_v t\,   . 
\end{equation}
Introducing the function $\omega$ such that
\begin{equation}
\omega= x+\lambda(u,v)\, t\, ,
\end{equation}
one rewrites the equations (\ref{Hodo2par}) as
\begin{equation}
t=\frac{\omega_u}{\lambda_u}\, , \qquad x = \omega -\frac{\lambda}{\lambda_u} \omega_u
\label{Hodo2par-sol}
\end{equation}
where $\omega$ obeys the equation
\begin{equation}
\omega_{uu}+\left( \lambda_v-\frac{\lambda_{uu}} {\lambda_u} \right)\omega_u-\lambda_u \omega_v=0\, .
\label{par-lineqn}
\end{equation}
Any solution of the equation (\ref{par-lineqn}) with given $\lambda(u,v)$ provides us via the relation (\ref{Hodo2par-sol}) a solution of the equations
(\ref{genparasys}).
On the other hand, for any solution $\omega$ of the equation (\ref{par-lineqn}) the formulae (\ref{Hodo2par-sol}) define a $\mathbb{R}^2 \to \mathbb{R}^2$ mapping
$(u,v) \to (t,x)$.  \par

The Jacobian $J$ for the mapping (\ref{Hodo2par-sol})  is
\begin{equation}
J=\left(\omega_v -\frac{\lambda_v}{\lambda_u} \omega_u\right)^2=t_u^2\, ,
\label{singcurgen2}
\end{equation}
and so the mappings (\ref{Hodo2par-sol})  are singular along the curve $\gamma$ given by
\begin{equation}
\sqrt{J} = \omega_v -\frac{\lambda_v}{\lambda_u} \omega_u =t_u=0\, .
\label{singcurgen}
\end{equation}
Note also that the parabolic mappings (\ref{Hodo2par-sol}), (\ref{par-lineqn}) the correspondence between infinitesimal areas in $(u,v)$ and $(t,x)$ planes 
\begin{equation}
\D t \wedge \D x = t_u^2 \D u \wedge \D v
\label{areap}
 \end{equation}
is of the second order near the curve $\gamma$. \par 

At each point of the curve $\gamma$ one has a tangent vector 
\begin{equation}
\vec{V} = (-t_{uv}, t_{uu})\Big{\vert}_\gamma\, .
\label{tan-sing}
\end{equation}
Following the Whitney's approach we introduce a vector field
\begin{equation}
\nabla_V = -t_{uv }\vert_ \gamma \partial_u+ t_{uu }\vert_ \gamma \partial_v= -t_{uv }\vert_ \gamma \partial_u+\lambda_u t_{v }\vert_ \gamma \partial_v\, .
\label{der-tan-sing}
\end{equation}
Its action on images of singular point $(u,v) \in \gamma$ is given by the formulae 
\begin{equation}
\begin{split}
\nabla_V (t,x)\vert_\gamma =& \lambda_u t_v^2 (1,-\lambda)\vert_\gamma \, , \\
\nabla_V^2 (t,x)\vert_\gamma =&  t_{v} \Big(
\lambda_{u} \left(t_{v}^2 
   \lambda_{uv}+2 t_{vv} t_{v} \lambda_{u}-2
   t_{uv}^2\right)-t_{v} t_{uv} \lambda_{uu}
\, , \\ %
&- \lambda_{u}  (t_{v}^2 \lambda_{u} \left(\lambda_{v} \lambda_{u}+2
   \lambda \lambda_{uv}\right)+t_{v} 
   (\lambda_{u} \left(2 t_{vv} \lambda 
   \lambda_{u}+t_{uv} \left(\lambda  \lambda_{v}-2 \lambda_{u}+1\right)\right)\\ & -3
   t_{uv} \lambda \lambda_{uu})+t_{uv}^2 \lambda \left(1-3
   \lambda_{u}\right) )
\Big) \vert_\gamma \, ,  
   \\
||\nabla_V^3 (t,x)\vert_\gamma|| =& P_3(t) t_v\vert_\gamma+Q_3(t) t_{uv}\vert_\gamma \, ,  \\ 
||\nabla_V^4 (t,x)\vert_\gamma|| = &P_4(t) t_v\vert_\gamma+Q_4(t) t_{uv}\vert_\gamma \, ,
\end{split}
\label{n-der-tan-sing}
\end{equation}
where $P_i$ and $Q_i$ are suitable polynomials in $t$ and their $u-$ and $v-$derivatives.
On the curve  $\gamma$ one has the relations
\begin{equation}
\begin{split}
& t_{uu}\vert_\gamma=\lambda_u t_{v}\vert_\gamma \, , \\
& t_{uuu}\vert_\gamma=(\lambda_{uu}-\lambda_u \lambda_v) t_{v}\vert_\gamma-\lambda_u t_{uv}\vert_\gamma\, , 
\end{split}
\end{equation}
and so on. \par

So, for the points on $\gamma$ for which $t_v \neq 0$ one has 
\begin{equation}
\nabla_V (t,x) \vert_\gamma \neq 0 \, ,  \qquad m=1,2,\dots\, ,
\label{m-der-tan}
\end{equation}
and, consequently, they should be folds according to \cite{W}. 
If for a point  (let us call it $0$) on $\gamma$ one has $t_v \vert_{0}=0$: then, according the first two equation of  (\ref{n-der-tan-sing}),
not only 
\begin{equation}
\nabla_V (t,x) \vert_0 = 0 \, ,
\label{1derpara0}
\end{equation}
but with necessity also 
\begin{equation}
\nabla_V^2 (t,x) \vert_0 = 0 \, .
\label{2derpara0}
\end{equation}
In this case $\nabla_V^m (t,x) \vert_\gamma \neq 0 $ for $m=3,4,\dots$ . So for the parabolic mappings
 (\ref{Hodo2par-sol}), (\ref{par-lineqn})  the equations (\ref{1derpara0}) and (\ref{2derpara0})
are not independent. \par 

This is a crucial difference  between the parabolic mappings  (\ref{Hodo2par-sol}), (\ref{par-lineqn})
and generic plane into plane mappings considered in \cite{W}. First for parabolic mappings  (\ref{Hodo2par-sol}), (\ref{par-lineqn}) there are no standard cusp points characterized by condition (\ref{cuspcond}) \cite{W}, i.e. by 
\begin{equation}
\nabla_V (t,x) \vert_0 = 0 \, , \qquad \nabla_V^2 (t,x) \vert_0 \neq 0\,  .
\label{cusp-cond-2}
\end{equation}
Second point is that for parabolic mappings only two of equations (\ref{3Wcond}), i.e.
\begin{equation}
J=0\, , \qquad \nabla_V f=0\, , \qquad \nabla_V^2 f=0\,
\end{equation}
characterizing bad set  (see \cite{W}) are independent.
So the bad set has codimension (defect) $2$ and condition $\delta>3$ for the Theorem 11A  in \cite{W} (p.$386$) is not verified. Consequently, the theorem 
13A \cite{W} on the existence of excellent mapping near any $\mathbb{R}^2 \to \mathbb{R}^2$ mapping is not applicable for parabolic mappings   (\ref{Hodo2par-sol}), (\ref{par-lineqn}). \par

Moreover, for the points on $\gamma$ at which $t_v\vert_{0}=0$ and $t_{uv}\vert_{0}=0$ not only 
\begin{equation}
\nabla_V (t,x) \vert_0 = 0 \, , \qquad \nabla_V^2 (t,x) \vert_0 = 0\, , \qquad  \nabla_V^3 (t,x) \vert_0 = 0 \, ,
\label{deeppara}
\end{equation}
but also all $\nabla_V^m (t,x) \vert_0 = 0$ for $m = 4,5,\dots$. In these points tangent vector for the curve $\gamma$  vanishes $\vec{V}=(0,0)$
and vector  field $\nabla_V$ is zero too. Such point are no more good points according to \cite{W}. It is noted that these singular points on $\gamma$ of higher order are not removable for  the family of mappings 
(\ref{Hodo2par-sol}), (\ref{par-lineqn}). \par
 
Finally we note that Whitney's choice of coordinate system given by (\ref{Wnorm}) is obviously not admissible for the parabolic mappings described by the equations 
(\ref{Hodoeqn-2par}). \par

These observations clearly indicate that in order to analyze singularities of the mappings (\ref{Hodo2par-sol}), (\ref{par-lineqn}) one should proceed in a different way.

\section{Hierarchy of singularities}
\label{Jorsing-sec}
Here we present a method to deal with singularities of mappings (\ref{Hodo2par-sol}), (\ref{par-lineqn}) different from the Whitney's original one. For convenience 
we will consider the simplest case with $\lambda=u$, i.e. the $\mathbb{R}^2 \to \mathbb{R}^2$ mappings governed by the Jordan system
  \begin{equation}
 \binom{u_t}{v_t}=\left( \begin{array}{cc}u & 1 \\ 0 &u \end{array} \right) \binom{u_x}{v_x}.
 \label{Jparasys}
\end{equation}
In the paper \cite{KK} it was shown that this system and its multi-component extensions are integrable, i.e. have infinite family of commuting symmetries. 
In hodograph plane the system (\ref{Jparasys}) is represented by equations 
\begin{equation}
x_u+ut_u=0\, , \qquad x_v +u t_v =t_u\, ,
\label{hodoJ}
\end{equation} 
and $t$ obeys the equation 
\begin{equation}
t_v=t_{uu}\, .
\label{heatJ}
\end{equation}
The second equation (\ref{hodoJ}) rewritten as $(x+ut)_v=t_u$ implies the existence of a function $W(u,v)$ such that
\begin{equation}
t=W_v\, , \qquad x+ut=W_u\, .
\label{imphodoJmap}
\end{equation}
Equation (\ref{heatJ}) means  that function $W$ obeys the equation  $W_v=W_{uu}$ (choosing the inessential integration ``constant'' to be zero).
So we have families of mappings $(u,v) \to (t,x)$ given by 
\begin{equation}
t=W_v\, , \qquad x= W_u-u W_v
\label{hodoJmap}
\end{equation}
 where $W$ is any solution of equation
 \begin{equation}
W_v=W_{uu}\, .
\label{potheatJ}
\end{equation}
In the form of equations for critical points of the function $W^*=xu+t\left( \frac{u^2}{2}+v\right) -W$
above equations has been considered in \cite{KK}. It was also shown that differential consequences of (\ref{hodoJmap}) and (\ref{potheatJ})   give
\begin{equation}
\label{der-sol-J}
\begin{split}
 u_x=&  \frac{W_{vv}}{(W_{uuu})^2}\, , \qquad v_x= -\frac{1}{W_{uuu}}\, ,  \\
  u_t=&  \frac{u\, W_{vv}}{(W_{uuu})^2}-\frac{1}{W_{uuu}}\, , 
 \qquad v_t= -\frac{u}{W_{uuu}} \, .
\end{split}
\end{equation}
and, hence, $u$ and $v$ are solutions of the system (\ref{Jparasys}) for any $W(u,v)$ obeying equation (\ref{potheatJ}).\par

Jacobian $J$ for the mappings (\ref{hodoJmap}), (\ref{potheatJ}) is
\begin{equation}
J=W_{uuu}^2\, ,
\label{JJ}
\end{equation}
i.e. (\ref{singcurgen2}) with $t_u=W_{uv}=W_{uuu}$ and $\omega=W_u$. So mappings (\ref{hodoJmap}), (\ref{potheatJ}) are singular on the curve $\gamma$  given by 
\begin{equation}
W_{uuu}=0\, .
\label{singJ}
\end{equation}
Formulas (\ref{der-sol-J}) clearly indicates that singularities  of the $\mathbb{R}^2 \to \mathbb{R}^2$ mappings (\ref{hodoJmap}), (\ref{potheatJ}) and gradient catastrophes for the system (\ref{Jparasys})
are intimately connected (for other systems see e.g. \cite{CF,Bers,Hor,GS,Rak1,Rak2,Giv,KS,Sul,Dub1,Dub2}).\par

In the case under consideration $t_v=W_{uuuu}$. So, for the point $\gamma(u_0,v_0)$ on $\gamma$ at which $W_{uuuu}\vert_{0}\neq 0$ one has
(formula (\ref{n-der-tan-sing}))
\begin{equation}
\nabla_V (t,x)\vert_{0}= W_{uuuu}^2 (1,-u)\vert_0\neq 0\, .
\label{der-t-s-J}
\end{equation}
So such point should be a fold according to \cite{W}. \par

 If at the point on $\gamma$  $W_{uuu}\vert_{0}=W_{uuuu}\vert_{0}=0$ and $W_{uuuuu}\vert_{0} \neq 0$ then  (formula (\ref{n-der-tan-sing}))
 \begin{equation}
 \nabla_V(t,x)\vert_{0}= \nabla_V^2(t,x)\vert_{0}= 0\, ,\qquad \nabla_V^2(t,x)\vert_{0}\neq 0\, .
 \end{equation}
 So such point is not a cusp, according Whitney's definition (\ref{cuspcond}). Further, if at a point $u_0,v_0$ on $\gamma$ 
 \begin{equation}
 \frac{\partial^3 W}{\partial u^3}\Big{\vert}_{0}= \frac{\partial^4 W}{\partial u^4}\Big{\vert}_{0}= \frac{\partial^5 W}{\partial u^5}\Big{\vert}_{0}=0\, , \qquad
  \frac{\partial^6 W}{\partial u^6}\Big{\vert}_{0}\neq 0\, ,
 \end{equation} 
 then according to the formulae (\ref{n-der-tan-sing}) one has
 \begin{equation}
 \nabla_V^m(t,x)\vert_0=0\, , \qquad m=1,2,3, \dots .
 \end{equation}
 These observations clearly show that the vector field $\nabla_V$ is not a right object to establish a gradation of singularities for mappings (\ref{hodoJmap}), (\ref{potheatJ})
 while the evaluation of the derivatives of $W$ with respect to $u$ at singular point seems to be a finer tool for that purpose. \par
 
 So we will use here a standard method of expansion near a point. In our case it is convenient to use particular  properties of equations (\ref{Jparasys})-(\ref{heatJ}).
 These equations are obviously invariant under Galilean transformation 
 \begin{equation}
 t\to t'=t\, , \qquad x \to x' = x+at\, , \qquad u \to u=u'-a\, , \qquad v \to v' = v-b\, ,
 \label{Galt}
 \end{equation} 
 and scaling transformations
 \begin{equation}
 u \to u'=\lambda u \, , \qquad v \to v' = \lambda^2 v\, ,
 \label{dilt}
 \end{equation}
 where $a,b$ and $\lambda$ are arbitrary parameters. Invariance under Galilean transformations means that for any mapping (\ref{hodoJmap}), (\ref{potheatJ})
 one can put the point  $(u_0,v_0)$
 at the origin without loss of generality. Different grading of $u$ and $v$ for the scaling transformation is preserved for small variations. 
 So for infinitesimal variations $\delta u$ and $\delta v$ one has
 \begin{equation}
 \delta u= \epsilon \overline{u}\, , \qquad  \delta v= \epsilon^2 \overline{v}\, ,
 \label{varu}
 \end{equation}
 where $\epsilon$ is a small parameter.
 Taking into account all that it is not difficult to show that expansion of function $W(u,v)$, obeying equation (\ref{potheatJ}), near the origin is of the form 
 \begin{equation}
 W(u,v)=\sum_{j\geq 0} \epsilon^j A_j P_j(\overline{u},\overline{v}) 
 \label{Wser}
 \end{equation}
 where $A_k= \partial^k W / \partial u^k \vert_0$ and $P_k(\ou,\ov)$ are elementary Schur polynomials (ESP) of two variables defined 
 by the generating relation 
 \begin{equation}
 \exp(z \ou+z^2 \ov) = \sum_{k\geq 0} z^j  P_j(\overline{u},\overline{v}) \, .
 \label{sciurg}
 \end{equation}
 One also has the following expansions
 \begin{equation}
 W_u= \sum_{j\geq 0} \epsilon^j   A_{j+1} P_j(\overline{u},\overline{v})\, ,  \qquad W_v=W_{uu}= \sum_{j\geq 0} \epsilon^j   A_{j+2} P_j(\overline{u},\overline{v})\, .
 \label{dW}
 \end{equation}
 At the plane $(t,x)$  the standard multi-scaling expansion near the point $(t_0,x_0)$ is given by (see e.g. \cite{DGZJ})
 \begin{equation}
 t=t_0 + \epsilon^\gamma \ot\, , \qquad  x=x_0 + \epsilon^\delta \ox\, ,
 \label{vart}
 \end{equation} 
 with parameters $\gamma$ and $\delta$ to be determined from the balance of dominant (in $\epsilon$) terms in both parts of the relations (\ref{hodoJmap}),  i.e.
 \begin{equation}
 t_0 + \epsilon^\gamma \ot = \sum_{j\geq 0} \epsilon^j   A_{j+2} P_j(\overline{u},\overline{v})\, ,
 \label{gJt}
 \end{equation}
 and
\begin{equation}
 x_0 + \epsilon^\delta \ox = \sum_{j\geq 0} \epsilon^j   A_{j+1} P_j(\overline{u},\overline{v})- \ou \sum_{j\geq 0} \epsilon^{j+1}   A_{j+2} P_j(\overline{u},\overline{v}) \, . 
 \label{gJx}
 \end{equation}
At the regular point $A_3 \neq 0$  and, since $t_0 = A_2$ and $x_0 = A_1$, one gets
\begin{equation}
\epsilon^\gamma \ot = \epsilon A_3 P_1 + o(\epsilon)\, , \qquad
\epsilon^\delta \ox = \epsilon^2 A_3 (P_2-\ou P_1) + o(\epsilon^2)\, .
\end{equation}
So $\gamma=1$, $\delta=2$ and at the leading order we have
\begin{equation}
\ot=\ou\, , \qquad \ox=-\frac{1}{2}\ou^2+\ov\, .
\label{regJ}
\end{equation}
This mapping is regular: Jacobian $J=1$ and change of variables $\ou=u^*$, $\ov=\frac{1}{2}{u^*}^2+v^*$ transforms it into
\begin{equation}
\ot=u^* \, , \qquad \ox=v^* \, .
\label{regJ-norm}
\end{equation}

On singular line (\ref{singJ}), $A_3=0$, but, in general, $A_4 \neq 0$. In this case formulae (\ref{gJt}), (\ref{gJx}) imply that
\begin{equation}
\epsilon^\gamma \ot = \epsilon^2  A_4 P_2 + o(\epsilon^2)\, , \qquad
\epsilon^\delta \ox = \epsilon^3 A_4 (P_3-\ou P_2) + o(\epsilon^3)\, .
\end{equation}
So $\gamma=2$, $\delta=3$ and
\begin{equation}
\ot= A_4 \left(  \frac{1}{2}\ou^2+\ov \right)\, , \qquad \ox=-A_4 \frac{\ou^3}{3} \, . 
\label{firstS3}
\end{equation}
In variables $u^*=\ou$ and $v^*= \ou^2/2+\ov$ it becomes
\begin{equation}
\ot=A_4 v^*\, , \qquad \ox=-A_4 \frac{{u^*}^3}{3}
\label{flexJ}
\end{equation}
and remains singular on the line $\ou=u^*=0$. Such mapping singularity can be referred as the \emph{flex}. \par

\begin{figure}[t]
	\centering
	\includegraphics[scale=0.5]{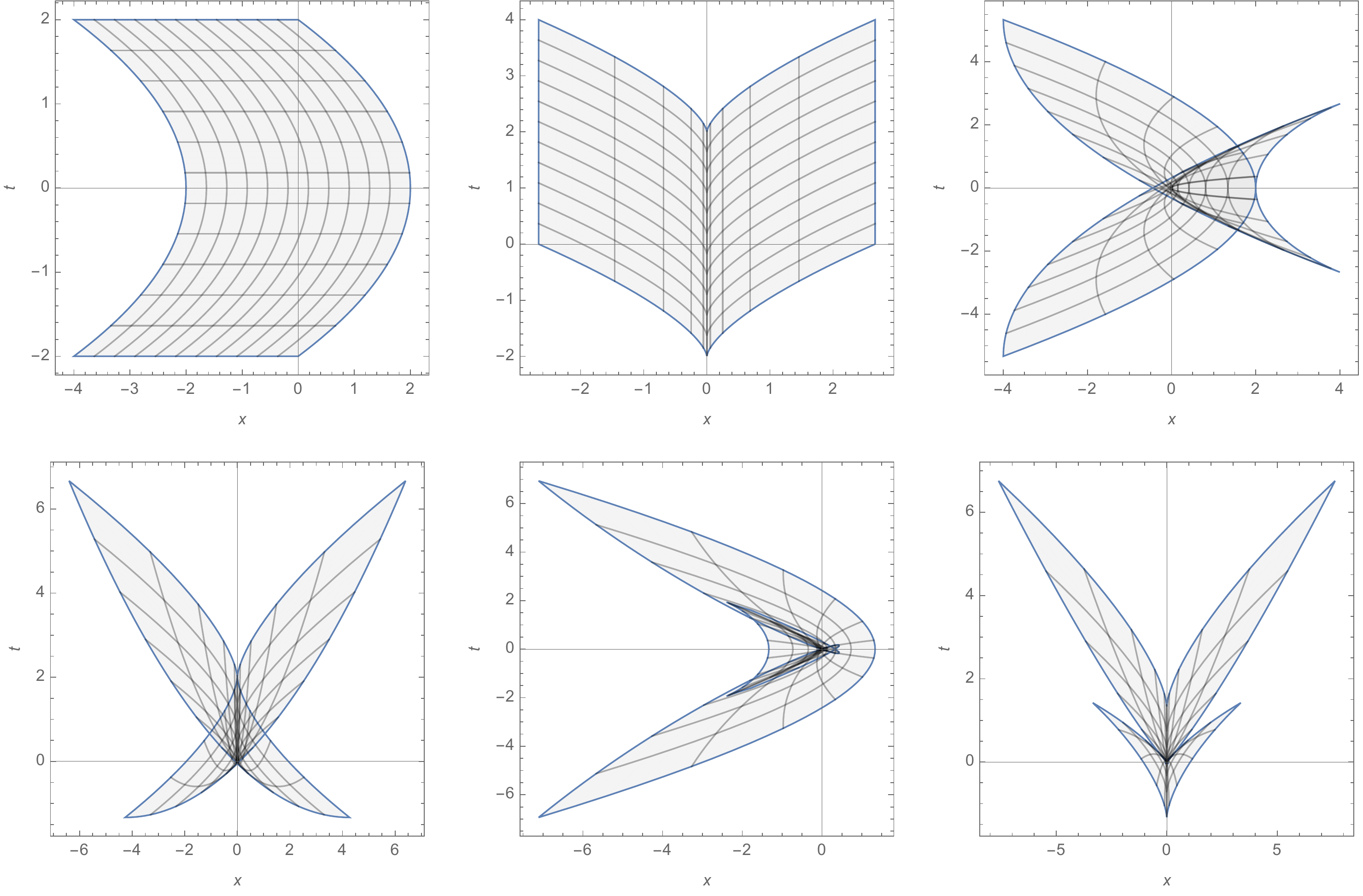}
	\caption{ Images in $(t,x)$-plane of the square $(-2,2) \times (-2,2)$ of the $(u,v)$-plane for the mappings (\ref{genJsing}) with
	$k=0,1,2$ (top) and $k=3,4,5$ (bottom) are shown. The top left plot for $k=0$ corresponds to a regular mapping.}
\label{singJ-fig}
\end{figure}

If at the point $\partial_u^3 W\vert_{0}=\partial_u^4 W\vert_{0}=0$, $\partial_u^5 W \vert_{0} \neq 0$, i.e. $A_3=A_4=0$, $A_5\neq 0$
one has $\gamma=3$ and $\delta=4$ and mapping near this point is given by 
\begin{equation}
\ot= A_5 P_3= A_5 \left( \frac{1}{6} \ou^3 + \ou\, \ov \right)\, , \qquad
\ox= A_5 (P_4-\ou P_3)= A_5 \left(  -\frac{1}{8} \ou^4- \frac{1}{2}  \ou^2 \ov + \frac{1}{2 }  \ov^2\right)\, . 
\label{cuspJ}
\end{equation}
It is singular  on the parabola $\pi \equiv \ou^2 /2 + \ov=0$ and the image of this parabola under the mapping (\ref{cuspJ}) is the $(3,4)$ curve
\begin{equation}
\ot \vert_{\pi}= -\frac{1}{3} A_5 \ou^3\, , \qquad
\ox \vert_{\pi}= \frac{1}{4} A_5 \ou^4\, .
\label{cuspJ34}
\end{equation}
These and higher order singularities are characterized by general condition
\begin{equation}
\frac{\partial^l W}{\partial u^l}\Big{\vert}_{0}=A_{l}=0\, , \quad l=1,2,\dots k+2\, , \qquad 
\frac{\partial^{k+3} W}{\partial u^{k+3}}\Big{\vert}_{0} = A_{k+3}\neq 0\, , \quad k=1,2,3,\dots\, .
\label{gen0J}
\end{equation}
In this case one has $\gamma=k+1$, $\delta=k+2$ and the mappings  near such points are (after rescaling $\ot$ and $\ox$ by $A_{k+3}$ )
\begin{equation}
\ot = P_{k+1}(\ou,\ov)\, , \qquad \ox = P_{k+2}(\ou,\ov)-\ou P_{k+1}(\ou,\ov)\, , \quad k=1,2,3,\dots
\label{genJsing}
\end{equation}
Thus, the parabolic mappings (\ref{hodoJmap}), (\ref{potheatJ}) have hierarchy of singularities  which corresponds to the gradation (\ref{gen0J}) and locally have the form (\ref{genJsing}). It is noted that the mappings (\ref{genJsing}) have multiplicity $k+2$.
 In the figure \ref{singJ-fig} we show the image of the mapping (\ref{hodoJmap})  near the lowest singular points. 
\par

Formulae (\ref{varu}) and (\ref{vart}) imply also that near singular points of order $k$ the infinitesimal areas in $(u,v)$ and $(t,x)$ planes are proportional to $\epsilon^3$
 and $\epsilon^{3+2k}$ namely
 \begin{equation}
 \delta u \wedge \delta v = \epsilon^3 \ou \wedge \ov\, ,\qquad \delta t \wedge \delta x= \epsilon^{3+2k} \ot \wedge \ox\, . 
 \end{equation}
 So, as $\epsilon \to 0$ 
 \begin{equation}
\delta t \wedge \delta x \sim \epsilon^{2k}  \delta u \wedge \delta v \, .
 \end{equation}
 Double rate decrease of area $(\sim (\epsilon^2)^k)$ near singular point is a characteristic feature of parabolic mapping (see also formula (\ref{areap})). \par
 
Finally, it is noted that higher singularities (\ref{gen0J}), (\ref{genJsing}) of mappings (\ref{hodoJmap})-(\ref{potheatJ}) are in one-to-one correspondence with higher order gradient catastrophes for the system (\ref{Jparasys}) (cf. \cite{KOpar}). 
\section{Structure of singular curves}
\label{singcurv-sec}
The expansion of $W_{uuu}$ near the point characterized by the conditions (\ref{gen0J}) is given by
\begin{equation}
W_{uuu} = \sum_{j\geq k} \epsilon^j A_{j+3} P_j{(\ou,\ov)}\, .
\end{equation}
Hence, near the $k$-th order singular point  the singular curve (\ref{singJ})
has locally the form
\begin{equation}
P_k{(\ou,\ov)} =0\, .
\label{zeroS}
\end{equation}
One gets the same result calculating the Jacobian $J_k$ of the
mappings (\ref{genJsing}), namely, $J_k=(P_k{(\ou,\ov)} )^2$.  \par

Due to the properties of ESPs, curves (\ref{zeroS}) are rather special. \par

Indeed one has the elementary
\begin{lem}
Elementary Schur polynomials have the following factorized form
\begin{equation}
P_{2n+1} (u,v) =\frac{1}{(2n+1)!}u \prod_{j=1}^n (u^2+4 \alpha_j^2 v)\, , \qquad
P_{2n} (u,v) =\frac{1}{(2n)!} \prod_{j=1}^n (u^2+4 \alpha_j^2 v)\, ,  
\label{S-Hroot}
\end{equation}
where $\alpha_j$ are roots of Hermite polynomials.
\end{lem}
{\bf Proof.}   
Due to the homogeneity of ESP it is obvious that 
\begin{equation}
P_j(u,v)= u^j P_j(1,y)\, , \qquad \mathrm{with} \quad y=\frac{v}{u^2}
\label{espsc}
\end{equation}
Polynomials $P_j(1,y)$ are defined by the generating relation
\begin{equation}
\exp(z+yz^2)= \sum_{j\geq 0} z^j P_j(1,y)\, .
\label{gnrtS1}
\end{equation}
Comparing (\ref{gnrtS1}) with the generating relation for the Hermite polynomials \cite{Sze}, i.e.
\begin{equation}
\exp(2 \alpha w-w^2)= \sum_{j\geq 0} \frac{w^j}{j!} H_j(1,y)\, ,
\label{gnrtH}
\end{equation}
one concludes that
\begin{equation}
P_j(1,y)= \frac{(-y)^{j/2}}{j!}H_j\left( \frac{1}{2\sqrt{-y}}\right)\, .
\end{equation}
Hence
\begin{equation}
P_j(u,v)= \frac{(-v)^{j/2}}{j!}H_j\left( \frac{u}{2\sqrt{-v}}\right)\, .
\label{HtoS}
\end{equation}
Since 
\begin{equation}
H_j(\alpha)= 2^j \prod_{i=1}^j  (\alpha-\alpha_i) \, ,
\end{equation}
where all roots of $H_j$ are distinct and real (\cite{Sze}),  one has the following factorized  form of ESP
\begin{equation}
P_j(u,v)=\frac{1}{j!} \prod_{i=1}^j (u-2 \alpha_i \sqrt{-v})\, .
\end{equation}
The roots $\alpha_i$ of Hermite polynomials are located symmetrically w.r.t. the origin \cite{Sze}. Consequently, one gets (\ref{S-Hroot}).  
{{$ \square$}} \par 

Formulae (\ref{S-Hroot}) imply that the singular curves (\ref{zeroS}) for mapping (\ref{gen0J}) are reducible, namely, they are unions of $n$ parabolas 
$u^2+4\alpha_i ^2 v=0$ for   $k=2n$ and unions of $n$ parabolas   $u^2+4\alpha_i ^2 v=0$  and the line $u=0$ for $k=2n+1$ where $\alpha_i$ are roots of 
Hermite polynomials. \par

It is easy to see that straight lines 
\begin{equation}
\ot \sim \ov^{n+1}\, , \qquad \ox= 0\, , \qquad k=2n+1
\end{equation}
are images of the line $\ou=0$. Images of parabolas are given by $(k+1,k+2)$ curves
\begin{equation}
\ot= B_k \ou^{k+1}\, , \qquad \ox= C_k \ou^{k+2}\, , \qquad k=2,3,4,\dots 
\label{sincurnew}
\end{equation}
with certain nonzero constants $B_k$ and $C_k$. 
The curves (\ref{sincurnew}) have singular points of orders $k$ at the origin $\ot=\ox=0$. The curvature $\kappa$ of these curves near the origin is unbounded since
\begin{equation}
\kappa \sim \ou^{-k}\, , \qquad \mathrm{as} \quad \ou \to 0.
\end{equation}
Tangent-normal pair for the curve  (\ref{sincurnew}) behaves smoothly for $k=2n$, $n=1,2,\dots$ while for $k=2n+1$, $n=1,2,\dots$  it exhibits an instant rotation
of angle $\pi$ passing the singular point $\ou=0$. The curves (\ref{sincurnew}) are $(2n+2,2n+3)$ cusps for $k=2n+1$.\par

\begin{figure}[t]
	\centering
	\includegraphics[scale=0.25]{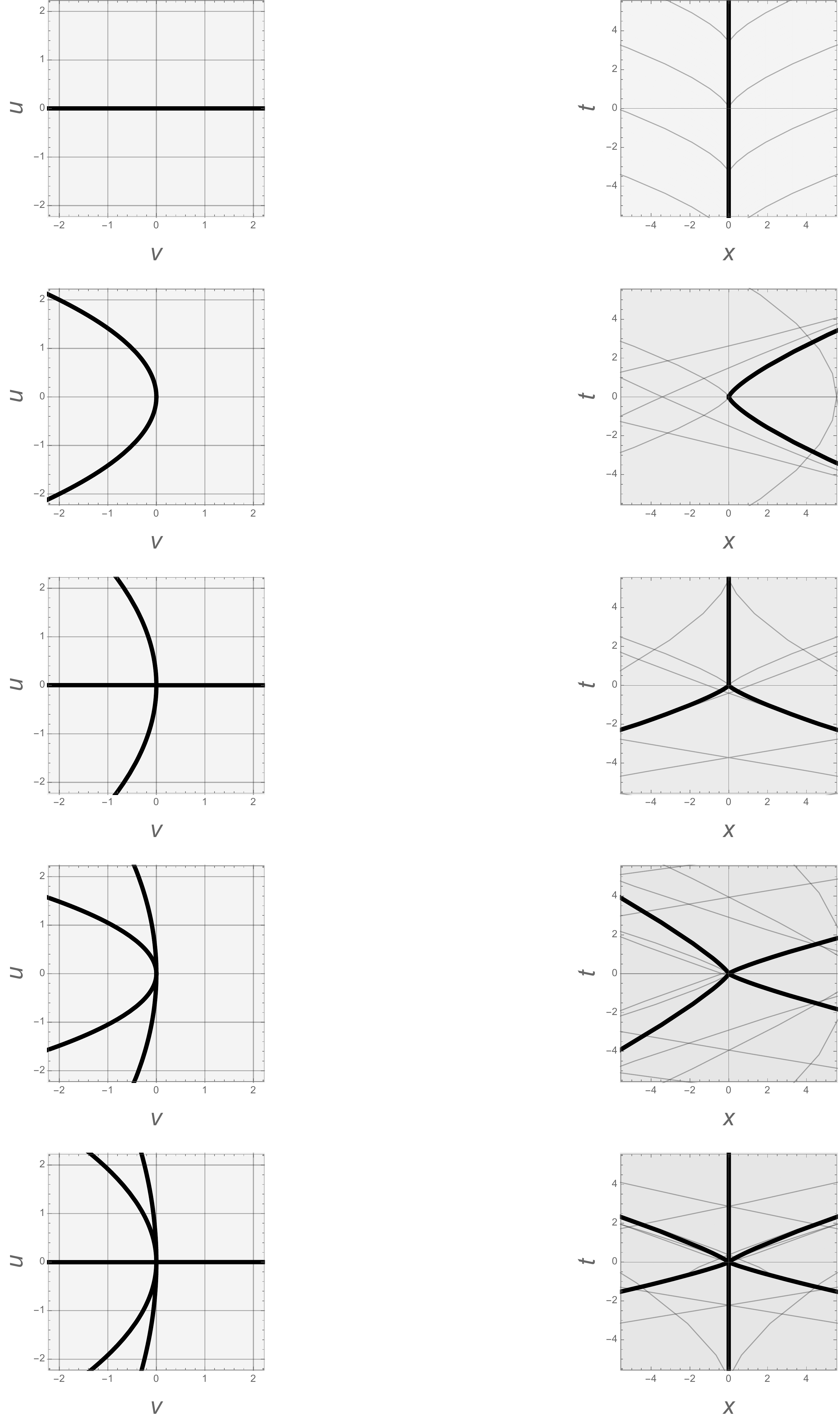}
	\caption{ Images in the $(t,x)$ plane (right column) of the singular curves mappings  (\ref{genJsing})
	of the singular curves in the $(u,v)$ plane (left column)  at $k=1,2,3,4,5$ are shown (from top to bottom).}
\label{singJ-curve-fig}
\end{figure}

In figure (\ref{singJ-curve-fig}) the images of singular curves are shown for $k=1,\dots,5$. \par

We would like to note that mappings (\ref{genJsing}) can be obtained also in a different, formal way. Indeed, let us consider a family of solutions of the equation 
(\ref{potheatJ}) of the form 
\begin{equation}
W(u,v;\tau_j)=\sum_{j \geq 0} \tau_j P_j(u,v)
\end{equation}
where $\tau_1, \tau_2, \tau_3,\dots$ are free parameters. The mapping (\ref{hodoJmap}) for such $W$ is given by
\begin{equation}
t= \sum_{j \geq 0} \tau_{j+2} P_j\, , \qquad x= \sum_{j \geq 0} \tau_{j+1} (P_j-j P_{j-1}) \, .
\label{multitimesW}
\end{equation}
Then the Jacobian $J=(\sum_{j \geq 0} \tau_{j+3} P_j )^2$ and singular curves are defined by the equations
\begin{equation}
\sum_{j \geq 0} \tau_{j+3} P_j(u,v) =0  \, .
\label{multitimessing}
\end{equation}
The equations (\ref{gen0J}) (for $u=v=0$) imply that
\begin{equation}
\tau_3=\tau_4=\dots=\tau_{k+2}=0\, , \qquad \tau_{k+3} \neq 0\, . 
\end{equation}
In such case the mapping (\ref{multitimesW}) takes the form 
\begin{equation}
t-\tau_2= \sum_{n\geq 0} \tau _{k+3+j} P_{k+1+j}\, , \qquad
x-\tau_1= \sum_{n\geq 0} \tau _{k+3+j} (P_{k+2+j}-u P_{k+1+j})\, ,
\label{multi-t-map}
\end{equation}
while the equation (\ref{multitimessing}) becomes
\begin{equation}
\sum_{n \geq 0 } \tau_{k+3+n}P_{k+1+n}(u,v)=0\, .
\label{multi-t-J}
\end{equation}
So the $k$-th order singular points of the mappings (\ref{multitimesW}) correspond to subspaces of codimension $k$ in the space with coordinates 
$(u,v,\tau_1,\tau_2,\dots)$. \par

The mapping (\ref{genJsing}) and singular curve (\ref{zeroS}) coincide with the  leading order terms $(n=0)$ in (\ref{multi-t-map}) and (\ref{multi-t-J}) 
for small $u$ and $v$ ($\ot=t - \tau_2$, $\ox=x - \tau_1$).
\section{Surface into the plane mappings: regularization}
\label{regul-sec}
In the paper \cite{KOpar} it was shown that the regularization of higher order gradient catastrophes for the system (\ref{Jparasys})  is achieved by embedding it 
into the multicomponent Jordan system 
\begin{equation}
\left( \begin{array}{c}
u_1\\ u_2 \\ \vdots \\ u_N
\end{array}
\right)_t=
\left( \begin{array}{ccccc}
u_1 & 1& 0 &\dots & 0 \\
0&u_1 & 1 &\dots & 0 \\
0& 0& u_1  &\dots & 0 \\
0 & 0& 0 &\dots & 1 \\
0 & 0& 0 &\dots & u_1 \\
\end{array}
\right)
\left( \begin{array}{c}
u_1\\ u_2 \\ \vdots \\ u_N
\end{array}
\right)_x \, .
\label{multiJsys}
\end{equation}
Hodograph equations for this system are given by the system \cite{KK} (see also \cite{KOpar})
\begin{equation}
x+u_1 t = \frac{\partial W^{(N)}}{\partial u_1}\, , \qquad t = \frac{\partial W^{(N)}}{\partial u_2}\, , \qquad  0 = \frac{\partial W^{(N)}}{\partial u_l}\, ,\quad l=3,4,\dots,N\, ,
\label{hodomultiJ}
\end{equation}
where function $W^{(N)}$ is a solution of the equations
\begin{equation}
\frac{\partial W^{(N)}}{\partial u_m} = \frac{\partial^m W^{(N)}}{\partial u_1^m}\, , \qquad m=1,2,\dots, N\, .
\label{heatmultiJ}
\end{equation}
One can view these relations   as the formulae which define mappings of subspace of the hodograph space $\mathbb{R}^N$ with coordinate $(u_1,u_2,\dots,u_N)$
into the plane $(t,x)$. Indeed, take a solution $W^{(N)}$ of the equations (\ref{heatmultiJ}). Then, last $N-2$ equations (\ref{hodomultiJ})  
\begin{equation}
 0 = \frac{\partial W^{(N)}}{\partial u_l}\, ,\quad l=3,4,\dots,N\, ,
 \label{surfdJ}
\end{equation}
define a surface $S^{(N)}(W^{(N)})$ in $\mathbb{R}^N$ and the first    two equations (\ref{hodomultiJ})
\begin{equation}
t = \frac{\partial W^{(N)}}{\partial u_2}\, , \qquad x= \frac{\partial W^{(N)}}{\partial u_1}-u_1 \frac{\partial W^{(N)}}{\partial u_2} \, ,
\label{multiJmap}
\end{equation}
define a mapping $S^N(W^{(N)}) \to (t,x)$-plane. It is noted that both the mapping (\ref{multiJmap}) and the surface $S^{(N)}(u_1,u_2,\dots u_N)$ 
are different for different functions $W^{(N)}(u_1,u_2,\dots u_N)$. Note also that on the subspace $u_3=u_4=\dots=u_N=0$ the system (\ref{multiJsys})
and mapping(\ref{multiJmap})
are reduced to the two-component Jordan system (\ref{Jparasys}) and mapping (\ref{hodoJmap}). It is readily seen, for instance, for solutions of the system 
(\ref{heatmultiJ}) in the form
\begin{equation} 
W^{(N)}(u_1,u_2,\dots u_N)=\int \D \lambda f(\lambda)  \exp(\lambda u_1+\lambda^2 u_2+\dots \lambda^N u_N)\, ,
\label{PotnJ}
\end{equation}
where $f(\lambda)$ is an arbitrary function. \par

Mapping (\ref{multiJmap}) is regular at the points where $\partial^{N+1} W^{(N)}/ \partial u_1^N \neq 0$. Indeed, the matrix
\begin{equation}
J_N = \left( 
{
\begin{array}{ccccc}
\displaystyle{\frac{\partial t}{ \partial u_1} } & \displaystyle{ \frac{\partial t}{ \partial u_2}}  & \dots  &
 \displaystyle{ \frac{\partial t}{ \partial u_{N-1}}}  &  \displaystyle{ \frac{\partial t}{ \partial u_N}}  \\
&&& \\
\displaystyle{\frac{\partial x}{ \partial u_1} } & \displaystyle{ \frac{\partial x}{ \partial u_2}}  & \dots  &
 \displaystyle{ \frac{\partial x}{ \partial u_{N-1}}}  &  \displaystyle{ \frac{\partial x}{ \partial u_N}} 
\end{array}
}
\right)
\end{equation}
due to the relations (\ref{heatmultiJ}) and (\ref{surfdJ}) is of the form
\begin{equation}
J_N = \left( 
\displaystyle{
\begin{array}{ccccc}
0&  \dots  & 0 &
 \displaystyle{ \frac{\partial t}{ \partial u_{N-1}}}  &  \displaystyle{ \frac{\partial t}{ \partial u_N}}  \\
&&& \\
0&  \dots  & 0 &
 \displaystyle{ \frac{\partial x}{ \partial u_{N-1}}}  &  \displaystyle{ \frac{\partial x}{ \partial u_N}} 
\end{array}
}
\right)
\label{matsurfN}
\end{equation}
In virtue of (\ref{heatmultiJ})-(\ref{multiJmap})  one has
\begin{equation}
\left\vert 
\begin{array}{cc}
 \displaystyle{ \frac{\partial t}{ \partial u_{N-1}}}  &  \displaystyle{ \frac{\partial t}{ \partial u_N}}  \\
& \\
 \displaystyle{ \frac{\partial x}{ \partial u_{N-1}}}  &  \displaystyle{ \frac{\partial x}{ \partial u_N}} 
\end{array}
\right\vert= \left(\frac{\partial^{N+1} W^{(N)}}{ \partial u_1^{N+1}}
\right)^2\, .
\label{detJNmap}
\end{equation}
So, in this case the rank of the matrix $J_N$ is $2$ and the mapping (\ref{multiJmap}) is regular (see e.g. \cite{Arn1}). \par

Let us now compare the mappings (\ref{multiJmap}) and those given by (\ref{hodoJmap}),(\ref{potheatJ}). At N=3 a family of functions $W^{(3)}(u_1,u_2,u_3)$
is formed by solutions of the system 
\begin{equation}
\frac{\partial W^{(3)}}{ \partial u_{2}}=\frac{\partial^2 W^{(3)}}{ \partial u_{1}^2}\, , \qquad
\frac{\partial W^{(3)}}{ \partial u_{3}}=\frac{\partial^3 W^{(3)}}{ \partial u_{1}^3}\, ,
\label{pot3J}
\end{equation}
 e.g. in the form (\ref{PotnJ}). Surface $S^{(3)}$ for given $W^{(3)}$ is defined by the equation
 \begin{equation}
 \frac{\partial^3 W^{(3)}}{ \partial u_{1}^3}=0\, ,
 \label{S3surf}
 \end{equation}
 and the mapping $S^{(3)} \to (t,x)$ is
 \begin{equation}
t=  \frac{\partial W^{(3)}}{ \partial u_{2}}\, , \qquad x=  \frac{\partial W^{(3)}}{ \partial u_{1}}-u_1 \frac{\partial W^{(3)}}{ \partial u_{2}} \, .
\label{red2map}
 \end{equation}
It is regular if on the surface (\ref{S3surf}) 
 \begin{equation}
 \frac{\partial^4 W^{(3)}}{ \partial u_{1}^4} \neq 0\, .
 \label{S3reg}
 \end{equation}
The mapping (\ref{hodoJmap}), (\ref{potheatJ}) with $k=1$ singularity ($u=u_1$, $v=u_2$) is given by 
 \begin{equation}
t=  \frac{\partial W}{ \partial u_{2}}\, , \qquad x=  \frac{\partial W}{ \partial u_{1}}-u_1 \frac{\partial W}{ \partial u_{2}} \, ,  \qquad 
\frac{\partial W}{ \partial u_{1}^3} = 0\, , \quad \frac{\partial^4 W}{ \partial u_{1}^4} \neq 0\, . 
\label{S3red}
 \end{equation}
where $W$ is a solution of the equation $\partial W/ \partial u_2 = \partial^2 W/ \partial u_1^2$. It is readily seen that under the restriction to the plane
$u_3=0$ such that $W^{(3)}(u_1,u_2,0)=W(u_1,u_2)$ the equations (\ref{pot3J})-(\ref{S3reg}) are reduced to those (\ref{S3red}). Intersection of the surface $S^{(3)}$
with the plane $(u_1,u_2)$
is a singular curve $ \partial^3 W/ \partial u_1^3=0$ for the mapping (\ref{hodoJmap}) and rank of the matrix $J_3$ (\ref{matsurfN}) becomes $1$
since $\partial t / \partial u_3 = \partial x / \partial u_3 = 0$.
So, a regular mapping $S^{(3)} \to (t,x)$ is singular under the restriction  of $\mathbb{R}^3$ to the plane $(u_1,u_2)$. 
 \begin{figure}[t]
	\centering
	\includegraphics[scale=0.5]{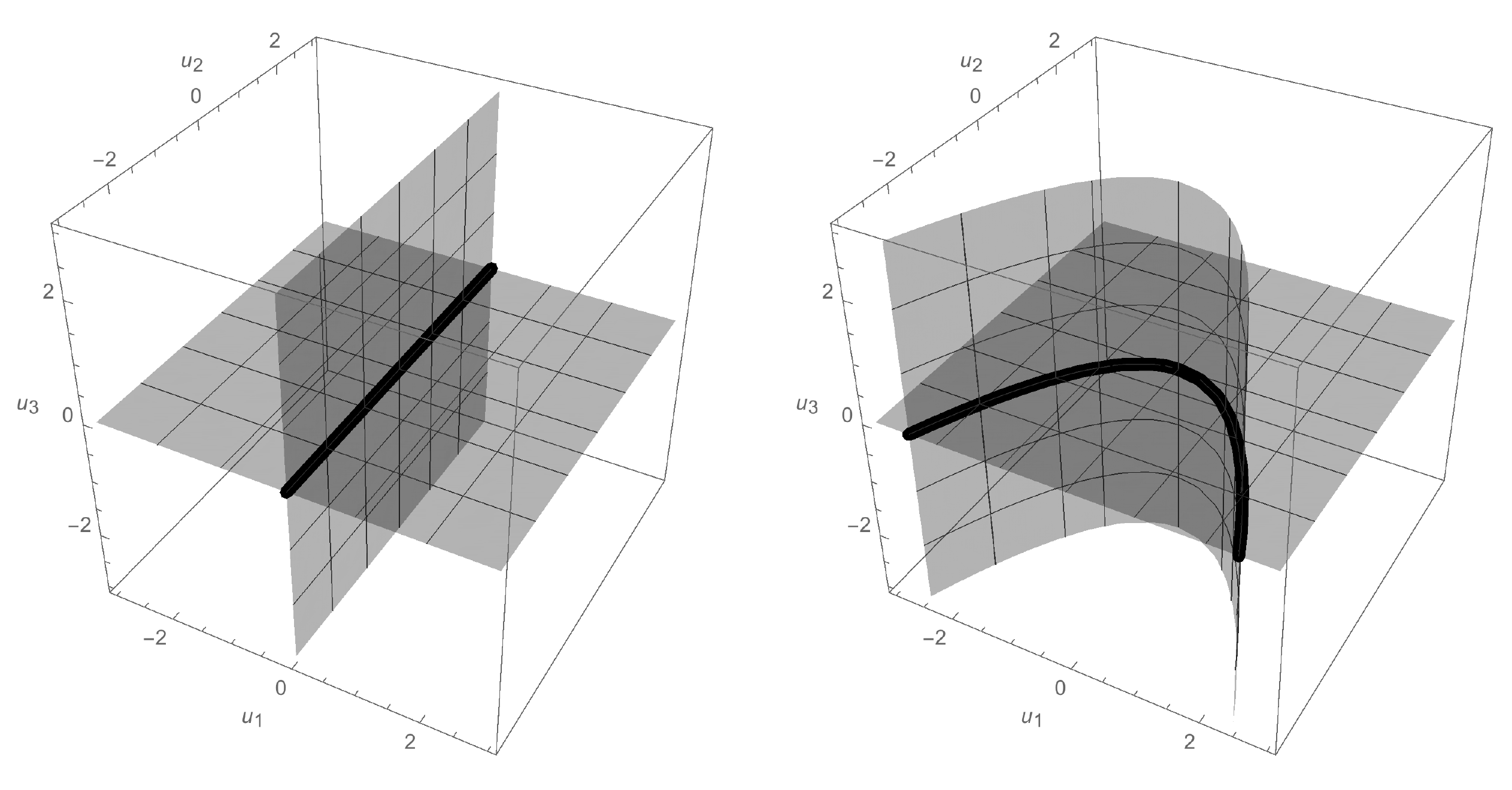}
	\caption{ 
	Regularizing interface $S^{(3)}$ for $k=1$ and the restriction of the hypersurface $\Gamma_4$ $(k=2)$ to the $(u_1,u_2,u_3)$
	space are shown. Vertical axis is the $u_3$ axis.
	}
\label{reg-surf-fig}
\end{figure}

 \begin{figure}[t]
	\centering
	\includegraphics[scale=0.6]{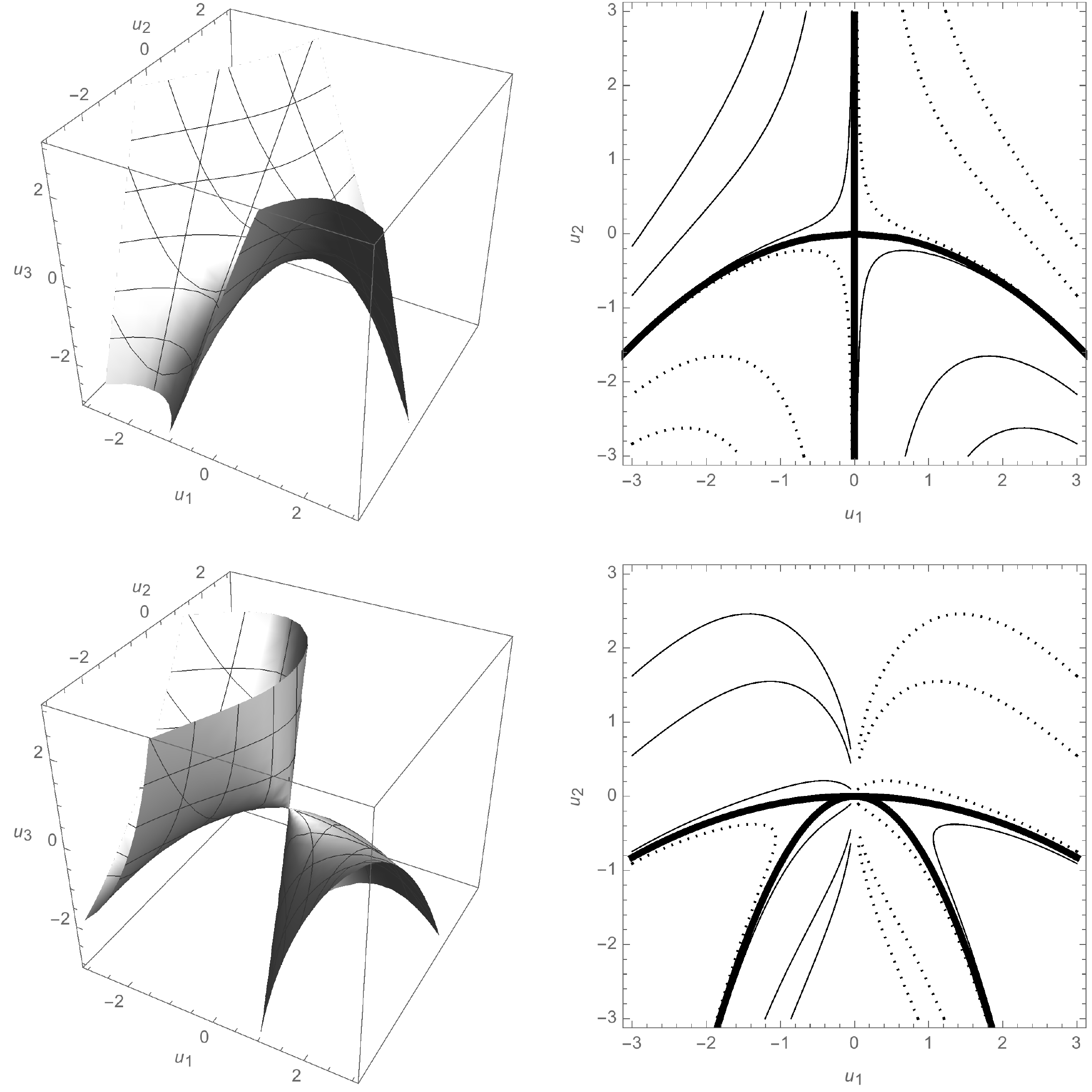}
	\caption{ Restriction of the hypersurfaces $\Gamma_5$ $(k=3)$ and $\Gamma_6$ $(k=4)$ to the  subspace with  coordinates $(u_1,u_2,u_3)$
	are shown. In the figures on the right the solid thick curve correspond to the level set $u_3=0$; the solid thin curves are  level sets $u_3= 0.1,2 ,4$
	and the dotted lines are  level sets $u_3= -0.1,-2 ,-4$.
	}
\label{Surf-slices-fig}
\end{figure}
 \par

This relation viewed in opposite
direction provides us with the regularization of the mappings (\ref{hodoJmap}), (\ref{potheatJ}). Indeed starting with the mapping (\ref{S3red}) for given functions $W(u_1,u_2)$ 
we first enlarge the space $(u_1,u_2)$  to $\mathbb{R}^3$ with coordinates $(u_1,u_2,u_3)$. Second we deform the function $W(u_1,u_2)$ to a function
$W^{(3)}(u_1,u_2,u_3)$, satisfying the equations (\ref{pot3J}) , such that 
$W^{(3)}(u_1,u_2,0)=W(u_1,u_2)$. Such deformation is, for example, the transition $N=2 \to N=3$ in the formula (\ref{PotnJ}) for the same $f(\lambda)$. 
Then one introduce the surface $S^{(3)}$ in $\mathbb{R}^3$ given by the equation (\ref{S3surf}) such that the condition (\ref{S3reg}) is satisfied and finally one define
the mapping $S^{(3)} \to (t,x)$-plane given by the formulae (\ref{red2map}) . 
In this deformation the $k=1$ singularity of the mapping (\ref{hodoJmap}) becomes a regular point for the mapping $S^{(3)} \to (t,x)$. \par

However, $k=2$ singularities  of the mappings (\ref{hodoJmap}), (\ref{potheatJ}) for which $\partial ^4 W/ \partial u_1^4 =0$ remain singularities of the mapping 
$S^{(3)} \to (t,x)$ (see (\ref{detJNmap}) at $N=3$). They can be regularized by extending the preimage plane $(u,v)$ to $\mathbb{R}^4$ with coordinates 
$u_1,u_2,u_3,u_4$ in the following way. Indeed, first one deforms a given function $W(u_1,u_2)$ to a function $W^{(4)}(u_1,u_2,u_3,u_4)$ obeying the equations
\begin{equation}
\frac{\partial W^{(4)}}{ \partial u_{2}}=\frac{\partial^2 W^{(4)}}{ \partial u_{1}^2}\, , \qquad
\frac{\partial W^{(4)}}{ \partial u_{3}}=\frac{\partial^3 W^{(4)}}{ \partial u_{1}^3}\, , \qquad
\frac{\partial W^{(4)}}{ \partial u_{4}}=\frac{\partial^3 W^{(4)}}{ \partial u_{1}^4}\, ,
\label{pot4J}
\end{equation}
and such that $W^{(4)}(u_1,u_2,0,0)=W(u_1,u_2)$ (see (\ref{PotnJ})).
Then one introduces a surface $S^{(4)}$ in $\mathbb{R}^4$ defined by the equations
\begin{equation}
 \frac{\partial^3 W^{(4)}}{ \partial u_{1}^3}=0\, , \qquad
  \frac{\partial^4 W^{(4)}}{ \partial u_{1}^4}=0\, , 
 \label{S4surf}
 \end{equation}
with $ {\partial^5 W^{(4)}}/{ \partial u_{1}^5}\neq0$ and consider the mapping  $S^{(4)} \to (t,x)$ given by
 \begin{equation}
t=  \frac{\partial W^{(4)}}{ \partial u_{2}}\, , \qquad x=  \frac{\partial W^{(4)}}{ \partial u_{1}}-u_1 \frac{\partial W^{(4)}}{ \partial u_{2}} \, .
\label{red3map}
 \end{equation}
 In this case the intersection of the hypersurface defined by the equation $\partial^3 W^{(4)}/\partial u_1^3=0$ with the plane $(u_1,u_2)$ coincides with the singular 
 curve for $k=2$. Intersection of the hypersurface  defined by the second equation (\ref{red2map}) gives another curve on the plane $(u_1,u_2)$. Intersection of 
 these two curves or, equivalently, intersection of the surface $S^{(4)}$ defined by both equations (\ref{S4surf}) with the plane $(u_1,u_2)$ is the $k=2$ 
 singular point of the mapping (\ref{hodoJmap}). \par
 
 The mapping (\ref{red3map}) is regular if $ {\partial^5 W^{(4)}}/{ \partial u_{1}^5}\neq0$. 
 So $k=2$ singularity of the mapping (\ref{hodoJmap})
 (given by (\ref{cuspJ34})) becomes regular point in a deformation of the mapping (\ref{hodoJmap}) into mapping  $S^{(4)}\to (t,x)$. 
 \par 
 
 In order to regularize higher singularities one should increase the dimension $N$. Comparing the formulae (\ref{hodomultiJ}), (\ref{heatmultiJ}) and 
 (\ref{hodoJmap})-(\ref{potheatJ}), (\ref{gJt})-(\ref{gJx}) and extending the argumentation presented in the above $N=3,4$ cases, one readily proves the following
 \begin{prop}
 The $k$-th order singularity of a mapping  (\ref{hodoJmap}),(\ref{potheatJ}) with a given function $W(u,v)$ is regularized by deformation of the mapping (\ref{hodoJmap}),(\ref{potheatJ}) into the mapping $S^{(2+k)}\to (t,x)$-plane.  In this deformation a function $W(u_1,u_2)$ is deformed 
 to a function $W^{(2+k)}(u_1,u_2,\dots, u_{k+2})$ which obeys to the equations
 \begin{equation}
\frac{\partial W^{(2+k)}}{ \partial u_{m}}=\frac{\partial^m W^{(2+k)}}{ \partial u_{1}^m}\, , \qquad m=2,3,\dots,2+k
\label{potnJ}
\end{equation}
 and satisfies the condition $W^{(2+k)}(u_1,u_2,0,\dots,0)=W(u_1,u_2)$. Surface $S^{(2+k)}$ in $\mathbb{R}^{2+k}$ is defined by the equations
 \begin{equation}
\frac{\partial^m W^{(2+k)}}{ \partial u_{1}^m}=0 \, , \qquad m=3,\dots,2+k
\label{Snsurf}
\end{equation}
 and the mapping $S^{(2+k)} \to (t,x)$ is given by 
  \begin{equation}
t=  \frac{\partial W^{(2+k)}}{ \partial u_{2}}\, , \qquad x=  \frac{\partial W^{(2+k)}}{ \partial u_{1}}-u_1 \frac{\partial W^{(2+k)}}{ \partial u_{2}} \, . 
\label{rednmap}
 \end{equation}
 \end{prop}
It is noted that such regularization is a local one. \par

In general an intersection of generic surface in $\mathbb{R}^{2+k}$ $(k \geq 3)$ with a plane $(u_1,u_2)$ is empty. However, in our case in virtue of the assumed 
compatibility of equations (\ref{gen0J}) a surface $S^{(2+k)}$ defined by (\ref{Snsurf}) intersects the $(u_1,u_2)$ plane at a point which is the $k$-th order 
singular point of the mapping (\ref{hodoJmap}). Namely, the first equation (\ref{Snsurf}), i.e. $\partial^3 W^{(2+k)}/\partial u_1^3 =0$, defines the hypersurface  in 
$\mathbb{R}^{2+k}$ intersection of which with the $(u_1,u_2)$-plane  coincides with the singular curve (\ref{singJ}) of the mapping (\ref{hodoJmap}). Intersection of 
other $k+1$ hypersurfaces defined by equations (\ref{Snsurf}) at $k=4,5,\dots,2+k$ with the $(u_1,u_2)$ plane gives us other $k+1$ curves. 
All these curves
by construction intersect the singular curve at a point of $k$-th order singularity of the mapping (\ref{hodoJmap}).\par

 Specifically, for the mappings (\ref{genJsing})
\begin{equation}
W^{(2+k)}(u_1, \dots, u_{2+k})=P_{3+k}(u_1, \dots, u_{2+k})
\label{pot2pk}
\end{equation}
where $P_l(u_1,\dots,u_n)$ are ESPs of $n$ variables defined by the generating relation
\begin{equation}
\exp \left(\sum_{l=1}^n z^l u_l  \right)= \sum_{l\geq 0} z^l P_l(u_1,\dots,u_n)\, .
\end{equation}
Then the mapping (\ref{rednmap}) takes the form 
\begin{equation}
t= P_{k+1}(u_1,\dots,u_{k+2})\, , \qquad x=P_{k+2}(u_1,\dots,u_{k+2})-u_1 P_{k+1}(u_1,\dots,u_{k+2})\, ,
\label{hmJ-part}
\end{equation}
while the surface $S^{(2+k)}$ is defined by equations
\begin{equation}
P_{3+k-m}(u_1,\dots,u_{2+k})=0\, ,  \qquad m=3,\dots, 2+k\, .
\label{genS2pk}
\end{equation}
Intersection of the hypersuperface $\Gamma_k: P_k(u_1,\dots,u_{2+k})=0$ with the $(u_1,u_2)$ plane is given by the curve $P_k(u_1,u_2,0,\dots,0)=0$ 
which is the singular curve (\ref{zeroS}). It is easy to show that the system of equations (\ref{genS2pk}) imply that $u_1=u_2=u_3=\dots=u_k=0$. So the $(u_{k+1},u_{k+2})$ plane is the regularizing surface 
$S^{(2+k)}$ in this case and the mapping (\ref{hmJ-part}) of this plane to the plane $(t,x)$ is of the form
\begin{equation}
t= P_{k+1}(0,\dots,0,u_{k+1},u_{k+2})\, , \qquad x=P_{k+2}(0,\dots,0,u_{k+1},u_{k+2})\, ,
\end{equation}
i.e.
\begin{equation}
t= u_{k+1}\, , \qquad x=u_{k+2}\, .
\end{equation}

First figure \ref{reg-surf-fig} (k=1) shows the regularizing surface $S^{(3)}$ for the mapping (\ref{firstS3}). It is the plane $u_1=0$ and
$W^{(3)}=u_1^4/24+ u_1^2u_2/2+u_2^2/2+u_1u_3 $.  
Second  figure \ref{reg-surf-fig} (k=2) shows the restriction of the hypersurface $\Gamma_4$ defined by the equation $u_1^2/2+u_2=0$ to the three-dimensional
space with coordinates $u_1,u_2,u_3$ $(u_4=0)$. Bold line represents the singular curve. Singular point  corresponds to $u_1=0$ on this curve. The $(u_3,u_4)$  
plane $(u_1=u_2=0)$ is the regularizing surface $S^{(4)}$ and $W^{(4)}={u_1^5}/{120}+ u_2 u_1^3/6+ u_3 u_1^2 /2 + u_2^2 u_1/2+u_4 u_1+u_2 u_3$\par

At the figure \ref{Surf-slices-fig} the restrictions of hypersurfaces $\Gamma_5$ ($k=3$) and $\Gamma_6$ ($k=4$) to the 3-dimensional
subspace with coordinates $(u_1,u_2,u_3)$ and corresponding  singular curves are shown.\par

Note that the above regularization of the mappings (\ref{genJsing}) formally can be viewed as the following simple procedure:
first, deform the ESPs $P_k(\ou,\ov)$ in (\ref{genJsing}), namely,
\begin{equation}
P_k(\ou,\ov) \to P_k(u_1,u_2,\dots,u_{k+2})
\end{equation}
and then restrict the deformed mapping (\ref{genJsing}) to the plane $u_1=u_2=\dots=u_k=0$.\par

One can follow similar procedure also in the case when $N<k+2$. Namely, first deform $P_k(\ou,\ov)$
\begin{equation}
P_k(\ou,\ov) \to P_k(u_1,u_2,\dots,u_{N})
\end{equation}
and then restrict the deformed mapping (\ref{genJsing}) to the plane $u_1=u_2=\dots=u_{N-2}=0$. One gets the mappings
\begin{equation}
t=P_{k+1}(0,\dots,0,u_{N-1},U_N)\, , \qquad x=P_{k+2}(0,\dots,0,u_{N-1},U_N)\, .
\label{mapredc2}
\end{equation} 
In contrast to the case $N=k+2$ all the mappings (\ref{mapredc2}) are singular. For instance, for $N=3$ and $k=2$ one has a fold
\begin{equation}
t=u_3\, , \qquad x=\frac{1}{2} u_2^2
\end{equation}
and for $N=3$ and $k=2$ one gets
\begin{equation}
t=\frac{1}{2} u_2^2\, , \qquad x=u_2u_3\, .
\end{equation}

For mappings (\ref{multitimessing}) the above results give a picture which arises if one considers the contribution of the dominant terms  (for small $u$ and $v$) only.
\section{On other parabolic type mappings}
\label{otherpara-sec}
Approach presented in previous sections can be applied to mappings governed by other parabolic systems of quasilinear PDEs connected with systems 
(\ref{genparasys}) and (\ref{multiJsys}). \par

First, the mappings (\ref{hodomultiJ})-(\ref{heatmultiJ}) are singular if $\partial^{N+1} W^{(N)}/\partial u_1^{N+1}=0$ (see (\ref{detJNmap})). Singular curve lies on the 
surface $S^{(2)}_N$ given by equations (\ref{surfdJ}) i.e.
\begin{equation}
\frac{\partial^l W^{(N)}}{\partial u_1^l}=0\, ,\qquad l=3,4,\dots, N\, ,
\label{surfSN}
\end{equation}
and defined by 
\begin{equation}
\frac{\partial^{N+1} W^{(N)}}{\partial u_1^{N+1}}=0\, .
\end{equation}
Higher singularities similar to $N=2$ case (\ref{heatJ}) are characterized by the condition
\begin{equation}
\frac{\partial^{l} W^{(N)}}{\partial u_1^{l}}=0\, , \quad  l=N+1, \dots, N+k \, ,\qquad 
\frac{\partial^{N+k+1} W^{(N)}}{\partial u_1^{N+k+1}}\neq 0\, , \quad k=1,2,3,\dots\, .
\label{gensing}
\end{equation}
At multiscaling expansion near a singular point of the order $k$ one has 
\begin{equation}
\delta u_n= \epsilon^n \ou_n\, , \quad k=1,2,\dots, N \, , \qquad 
\delta t= \epsilon^{N+k+1} \ot \, , \qquad 
\delta x= \epsilon^{N+k} \ox \, , 
\label{genexp}
\end{equation}
Performing this expansion for the mapping (\ref{multiJmap}), one gets 
\begin{equation}
\ot=A _{N+k+1} P_{N+k-1}(\ou)\, , \qquad
\ox=A _{N+k+1} (P_{N+k}(\ou)-\ou_1 P_{N+k-1}(\ou)  )\, ,
\label{genmap}
\end{equation}
where $A_m = \partial^m W^{(n)}/ \partial u_1^m \vert_0$  and $P_k(\ou_1,\dots, \ou_N)$ are ESPs with $N$ variables . \par

Since the expansion (\ref{genexp}) should be performed on the surface $S^{(2)}_N$ one, using the expansion
\begin{equation}
\frac{\partial^{l} W^{(N)}}{\partial u_1^{l}}= \frac{\partial^{l} W^{(N)}}{\partial u_{l}} \Big{\vert}_0+\sum_{m\geq 1} \epsilon^m A_{l+m} P_m(\ou)\, ,
\end{equation}
 gets from (\ref{surfSN}) the equations
\begin{equation}
P_{N+k+1-l}(\ou)=0\, ,\qquad l=3,4,\dots,N\, .
\label{zeroP}
\end{equation}
So, near to the $k$-th order singular point (\ref{gensing}) the mapping (\ref{multiJmap}) have the form (\ref{genmap}) and represent themselves the mappings
of the  surfaces $S^{(2)}_N$ defined by equations (\ref{zeroP}) into the plane $(\ot,\ox)$. \par

At $N=3$ surfaces $S^{(2)}$ are given by 
\begin{equation}
P_{k+1}(\ou_1,\ou_2,\ou_3)=0\, .
\label{Pk3}
\end{equation}
In particular for $k=1$ it is a cylindrical surface generated by the parabola
\begin{equation}
\frac{1}{2} \ou_1^2+\ou_2=0
\label{surcyl}
\end{equation}
and the mapping (\ref{genmap}) is
\begin{equation}
\ot=A_5\left( -\frac{1}{3} \ou_1^3 +\ou_3 \right)\, , \qquad \ox= \frac{A_5}{4} \ou_1^4\, .
\label{genfoldgen}
\end{equation}
It is a fold type singularity. In the variables $u_1^*=\ou_1$, $u_3^*=\ou_3 -{\ou_1^3}/{3}$ it assumes the form 
$ \ot=A_5 u_3^* \, ,  \ox= {A_5} {u_1^*}^4$. \par

Comparing (\ref{cuspJ34}) and (\ref{genfoldgen}), one sees that $(3,4)$ singularity for the mapping (\ref{hodoJmap}) is deformed into the fold singularity at 
$N=3$ and the singular curve $\ou/2+\ov=0$ at $N=2$ becomes a curve generating surface (\ref{surcyl}) in $\mathbb{R}^3. $\par 

At $k=2$ the surface $S^{(2)}_3$ is a cubic $\ou_1^3/6+\ou_1\ou_2+\ou_3=0$ and so on. Thus, $k$-th order singularity (\ref{genJsing}) at $N=2$ is deformed into the $k-1$-th order singularity (\ref{genmap}) for $N=3$, while the singular curve (\ref{zeroS}) is transformed into the surface $P_k(\ou_1,\ou_2,\ou_3)=0$ (\ref{Pk3}).\par

For $N=4$ the surface $S^{(2)}_4$ in $\mathbb{R}^4$ is given by the equations
\begin{equation}
P_{k+1}(\ou)=0\, ,\qquad P_{k+2}(\ou)=0\, .
\end{equation}
At $k=1$ it is a cylindrical surface defined by equations 
\begin{equation}
\ou_2+\frac{1}{2}\ou_1^2=0\, , \qquad \ou_3-\frac{1}{3}\ou_1^3=0\, .
\end{equation}
In general, it is readily seen that the $k$-th order singularity  (\ref{genJsing}) becomes the $k+1-N$-th order singularity (\ref{genmap}) under the deformation 
$N=2 \to N\neq 3$ while the conditions (\ref{gen0J}) are transformed into equation (\ref{zeroP}) defining a surface $S^{(2)}_N$. 
\par

We would like also to note that the singularities (\ref{genmap}), (\ref{zeroP}) are in one-to-one correspondence with the higher order gradient catastrophes fo the system (\ref{multiJsys}) \cite{KOpar}. \par

Second example is connected with the equations which describe higher symmetries of the system (\ref{genparasys}). They are given by (\cite{KK})
\begin{equation}
\left( \begin{array}{c}
u\\ v
\end{array}
\right)_{t_s}=
\left( \begin{array}{cc}
p_s & p_{s-1} \\
0& p_s
\end{array}
\right)
\left( \begin{array}{c}
u\\ v
\end{array}
\right)_x \, , \qquad s=2,3,\dots\, .
\label{2psys}
\end{equation}
For fixed $s$ the hodograph equations are \cite{KK}
\begin{equation}
t_s p_{s-1}(u,v)=W_v\, ,\qquad x+t_s p_s(u,v)=W_u\, ,
\end{equation}
which define the mapping $(u,v)\to (t_s,x)$ for a given $W$ obeying the equation $W_u=W_{vv}$. \par

Further, for the system (\ref{multiJsys}) higher symmetries are described by the system \cite{KK} 
\begin{equation}
\left( \begin{array}{c}
u_1\\ u_2 \\ \vdots \\ u_N
\end{array}
\right)_t=
\left( \begin{array}{ccccc}
p_{s} & p_{s-1}& 0 &\dots & p_{s-N+1} \\
0&p_{s} & p_{s-1} &\dots & 0 \\
0& 0& p_{s}  &\dots & 0 \\
0 & 0& 0 &\dots & 1 \\
0 & 0& 0 &\dots & p_{s} \\
\end{array}
\right)
\left( \begin{array}{c}
u_1\\ u_2 \\ \vdots \\ u_N
\end{array}
\right)_x \, .
\label{multipsys}
\end{equation}
where $p_s(u)$ are ESP polynomials of $N$-variables and $p_s=0$ at $s<0$. Corresponding hodograph equations \cite{KK} define mappings 
$S^{(2)}_N(W) \to (t,x)$ given by  
\begin{equation}
t_s p_{s-1}(u)=W_{u_2}^{(N)}\, ,\qquad x+t_s p_s(u)=W_{u_1}^{(N)}
\label{hodop}
\end{equation}
of surfaces $S^{(2)}_N$ in $\mathbb{R}^N$ defined by equations
\begin{equation}
\frac{\partial^l W^{(N)}}{\partial u_1^l}=0\, ,\qquad l=3,4,\dots,N \, ,
\label{Wpeqn}
\end{equation}
where $W^{(N)}$ obeys equations (\ref{heatmultiJ}). \par

Now, if one consider the family of $N-1$ commuting system (\ref{multipsys}) $(s=1,2,\dots,N-1)$, the hodograph equations for their common solutions 
$u_n(x,t_1,\dots,t_{n-1})$, $n=1,\dots,N$
instead of (\ref{hodop}), (\ref{Wpeqn}) assume the form \cite{KK}
\begin{equation}
\sum_{s=0}^{N-1} t_s P_{s+1-l}(u) = W_{u_l}^{(N)}\, , \qquad l=1,2,\dots,N\, ,
\end{equation} 
where $t_0=x$. This system of equation or their more explicit form
\begin{equation}
t_{N-1}=W_{u_N}^{(N)}\, ,\qquad
t_{N-2}+t_{N-1}P_1(u)=W_{u_N}^{(N)}\, ,\qquad
\sum_{s=0}^{N-1}t_{s} P_s(u)=W_{u_1}^{(N)}\, ,
\end{equation}
define mappings $(u_1, u_2,\dots, u_N)\to (t_0,t_1,\dots,t_{N-1})$ of $\mathbb{R}^N \to \mathbb{R}^N$ of parabolic type.\par

Finally we note that  since the parabolic systems (\ref{genparasys}), (\ref{multiJsys}), (\ref{multipsys}) can be viewed as the degeneration of strictly hyperbolic hydrodynamic type systems via certain confluences process \cite{KK} the parabolic mappings and their singularities considered in this paper can be 
recovered via certain degeneration of mappings (confluence type process)  and singularities  associated with strictly hyperbolic systems of PDES.
\section{Hyperbolic case}
\label{hyper-sec}
Singularities of solutions and of mappings for strictly hyperbolic two-component quasilinear systems of PDEs  have been studied in several papers
 (see e.g. \cite{Kri,Rak1,Rak2,KS,Sul,Dub2}). Standard folds and cusps are their generic singularities.
 Here we will briefly discuss such systems and their singularities in order to emphasize differences between hyperbolic and parabolic cases.\par 
 
 Let us consider the two component system written in terms of Riemann invariants $r$ and $s$ and characteristic speeds $R,S$, i.e.
 \begin{equation}
 r_t=R(r,s)r_x\, ,\qquad s_t=S(r,s)s_x\, .
 \label{2hypsys}
 \end{equation}
In hodograph space it is of the form 
 \begin{equation}
 x_r+S(r,s) t_r=0\, ,\qquad x_s+R(r,s)t_s =0\, .
 \label{2hyphodo}
 \end{equation} 
 and $t$ obeys the equation 
 \begin{equation}
 (S-R) t_{rs}= R_r t_s -  S_s t_r\, .
 \label{teqnhyp}
 \end{equation} 
 The Jacobian of the mapping $(r,s) \to (t,x)$ is
 \begin{equation}
 J=(S-R) t_r t_s\, .
 \label{Jhyp}
 \end{equation}
 The mapping $(r,s) \to (t,x)$ is singular if 
 \begin{equation}
 S=R\, ,
 \label{HEt}
 \end{equation}
 or\begin{equation}
t_r t_s=0 \, .
 \label{cathyp}
 \end{equation}
 First case (\ref{HEt}) is realized on the transition line between  hyperbolic and elliptic domains. Here we will consider  the generic second situation  
 (\ref{cathyp}), namely, the case
 \begin{equation}
 t_r=0, \qquad t_s \neq 0 \qquad \mathrm{or} \qquad  t_r \neq 0, \qquad t_s = 0\, .
 \end{equation} 
 Let the singular curve be given by $t_r=0$. The associated Whitney vector field $\nabla_V$ is
 \begin{equation}
 \nabla_V = -t_{rs}\vert_{t_r=0}\partial_r +t_{rr}\vert_{t_r=0}\partial_s= \frac{R_rt_s}{R-S}\Big{\vert}_{t_r=0}\partial_r +t_{rr}\vert_{t_r=0}\partial_s\, .
 \label{vechyp}
 \end{equation}
 Calculating $\nabla_V (t,x)$, one gets
 \begin{equation}
 \nabla_V (t,x)\vert_{t_r=0}=t_st_{rr} (1,-R)\vert_{t_r=0}\, .
 \label{vechyptx}
 \end{equation}
 If $t_rr \vert_{t_r=0} \neq 0 $ then $ \nabla_V (t,x)\vert_{t_r=0} \neq 0$. So the points where $t_r=0$, $t_s \neq 0$,  $t_{rr} \vert_{t_r=0} \neq 0 $ are folds
 according to Whitney's definition. In the case  $t_r=0$, $t_s \neq 0$,  $t_rr \vert_{t_r=0} = 0 $ one has
 \begin{equation}
 \nabla_V^2 (t,x)\vert_{t_r=0}=\frac{R_r t_s^2 t_{rrr}}{R-S} (1,-R)\vert_{t_r=t_{rr}=0}\, .
 \label{2vechyptx}
 \end{equation}
 Thus, at points where $t_r=t_{rr}=0$, but $t_{rrr}\vert_{t_r=0} \neq 0$, $t_s\neq 0$, one has  
 \begin{equation}
 \nabla_V (t,x)\vert_{t_r=0} =0 \, ,\qquad \nabla_V^2 (t,x)\vert_{t_r=0}\neq 0\, .
 \end{equation} 
 These points are cusps. \par
 One can show by induction that the gradation of vanishing $\nabla^k_V (t,x) $ corresponds to the gradation in derivatives $t_r,\, t_{rr},\, t_{rrr}, \dots$\, .
 So it is natural to introduce the gradation of singularities of mapping $(r,s) \to (t,x)$ according to a number $n$ of vanishing derivatives of $t$ w.r.t. $r$. 
 Namely, we refer to a singularity to be of the order $n$ if
 \begin{equation}
 \frac{\partial^k t}{ \partial r^k}\Big{\vert}_0 =0\, , \quad k=1,2,\dots, n \, , \qquad  \frac{\partial^{n+1} t}{ \partial r^{n+1}} \Big{\vert}_0 \neq 0\, , \qquad t_s\neq 0\, . 
 \label{hypncat}
 \end{equation}
In this case straightforward calculation gives
\begin{equation}
\nabla^k_V(t,x)\vert_{0} =0 \, , \qquad k=1,2, \dots n-1\, ,
\label{hypncat0}
 \end{equation}
 and 
\begin{equation}
\nabla^n_V(t,x)\vert_{0} = \frac{R \partial_r^{n-1} R t_s^n \partial_r^{n+1}t}{(R-S)^{n-1}}(1,-R)\Big{\vert}_{0}  \neq 0
\label{hypncatno0}
 \end{equation}
So, there is a one-to-one correspondence between the gradation in terms of vector fields $\nabla_V$ and in terms of derivative $\partial_r$. At points for which the conditions (\ref{hypncat0}), (\ref{hypncatno0}) are verified, the mappings $(r,s) \to (t,x)$ have $n-$th order singularities.
Such definition of higher order singularities is a natural extension of the original Whitney's definition of fold and cusps. All the results presented here can be applied 
in the case $t_s=0$, $t_r\neq 0$ with the exchange $r \leftrightarrow s$. 
We see that the original Whitney's approach is fully applicable in the hyperbolic case. \par

Finally we note that in the weakly-nonlinear case when $R_r=S_s=0$ all the formulae degenerate drastically. 
Indeed, in this case $t_{rs}=0$ and, hence, $t=a(r)+b(s)$ where
$a$ and $b$ are arbitrary functions.  Then in the case $t_r=0$, $t_s \neq 0$ it holds 
\begin{equation}
\nabla_V = a_{rr}\vert_{ a_r=0} \partial_s\, .
\end{equation}
So this vector fields is factorized  and also
\begin{equation}
\nabla_V^k (t,x) = a_{rr}^k\vert_{ a_r=0} (\partial_s^k t, \partial_s^k x) = 
a_{rr}^k\vert_{ a_r=0} \partial_s^{k-1} \Big( b_s  (1,  -R  ) \Big)  \, .
\end{equation}
Hyperbolic and elliptic cases will be analysed in more detail in subsequent publication.
\subsection*{Acknowledgments}
B.G.K. thanks Prof. W. Schief for useful suggestions on the paper.
This research has been supported by grant H2020-MSCA-RISE-2017 Project No. 778010 IPaDEGAN.
The author G.O. gratefully acknowledge the auspices of the GNFM Section of INdAM under which part of this work was carried out. 


\begin{thebibliography}{00}
\bibitem{Ale} G. Alessandrini, Critical points of solutions of elliptic equations in two variables,  Ann. Scuola Norm. Super. Pisa Cl. Sci. (4) 14 , 229-256 (1987)
\bibitem{Arn1} V. I. Arnold, S. M. Gusein-Zade and A. N. Varchenko,  Singularities of Differentiable Maps vol I-II Basel: Birkhäuser (1985-1988)
 \bibitem{Arn2}  V. I. Arnold,  Normal forms of functions in neighborhoods of degenerate critical points, Russ. Math. Surv. 29 10-50   (1974)  
\bibitem{Bers} L. Bers, Mathematical aspects of subsonic and transonic gas dynamics,  John Wiley \& Sons, New York (1958)
\bibitem{CFO}  {R. Camassa, G.  Falqui and G.  Ortenzi,}
Two-layer interfacial flows beyond the Boussinesq approximation: a Hamiltonian approach,
{ Nonlinearity} {30} 466-491 (2017)
\bibitem{CF} R. Courant and K.O. Friedrichs, Supersonic Flow and Shock Waves, Springer-Verlag, New York (1962)
\bibitem{DGZJ}  P. Di Francesco, P. Ginsparg and  J. Zinn-Justin, 2D gravity and random matrices, Physics Reports 254 (1-2) 1-133 (1995)
\bibitem{Dub1} B. A.  Dubrovin, {On Hamiltonian Perturbations of Hyperbolic
Systems of Conservation Laws, II: Universality of Critical Behaviour}, Commun. Math. Phys. 267 117-139 (2006)
\bibitem{Dub2} B. A.  Dubrovin, On universality of critical behaviour in Hamiltonian PDEs, 
	 Geometry, topology, and mathematical physics : S.P. Novikov's seminar : 2006-2007 / 
	V.M. Buchstaber, I.M. Krichever, editors. - Providence, R.I. : American Mathematical Society  59-109 (2008) 
\bibitem{DGKM}  B. Dubrovin, T.  Grava, C.  Klein and A. Moro, On critical behaviour in systems of Hamiltonian
partial differential equations, J. Nonlinear Sci. 25 631-707 (2015)
\bibitem{Giv} A. B. Givental, Whitney singularities of solutions of partial differential equations, J. Geom. Phys.
15(4) 353-368 (1995)
\bibitem{Gol} M. Golubitsky and V. Guillemin, Stable Mappings and Their Singularities, Graduate Texts in Mathematics (14) Springer-Verlag (1973)
\bibitem{Guc} J. Guckenheimer, Catastrophes and partial differential equations, Ann. Inst. Fourier (Grenoble) 23, 31-59 (1973)
\bibitem{GS} A. Gurevich and A. Shvartsburg,  Exact solutions of the equations of nonlinear geometric optics, Soviet Physics JETP 31 (6) 1084-1089 (1970)  
\bibitem{Hor} L. V. H\"ormander, On the singularities of solutions of partial differential equations, Matematika, 16:6  33-59 (1972) 
\bibitem{KMAM} B. G. Konopelchenko, L.  Mart\'inez Alonso, E. Medina, Hodograph solutions of the dispersionless coupled KdV hierarchies, critical points and the 
Euler-Poisson-Darboux equation, Journal of Physics A: Mathematical and Theoretical 43(43) 434020 (2010)
\bibitem{KK} B. G. Konopelchenko and Y. Kodama, Confluence of hypergeometric functions and integrable hydrodynamic-type systems, Theor. Math. Phys. 188 429-55
(2016)
\bibitem{KOSapm} B. G. Konopelchenko and G. Ortenzi, Quasi-Classical Approximation in Vortex Filament Dynamics, Integrable Systems, Gradient Catastrophe, and Flutter, Stud. Appl. Math. 130:2 167-199  (2013) 
\bibitem{KOJor} B. G. Konopelchenko and G. Ortenzi, {Jordan form, parabolicity and other features of change of type transition for hydrodynamic type systems},
J. Phys. A: Math. Theor. 50 (2017) 215205 (22pp)
\bibitem{KOpar} B G Konopelchenko and G Ortenzi, Parabolic regularization of the gradient catastrophes for the Burgers-Hopf equation and Jordan chain,
  J. Phys. A: Math. Theor. 51 275201 (2018)
\bibitem{Kri} A. P. Krishchenko, The structure of singularities of the solutions of quasilinear equations, Uspekhi Mat. Nauk. 31 219-220 (1976) 
\bibitem{KS} V. R. Kudashev, and B. I. Suleimanov, Characteristic features of some typical spontaneous intensity collapse processes in unstable media, JETP Letters 
62(4) 358-363 (1995)
\bibitem{L-VI} L. D. Landau, Fluid Mechanics, Pergamon press (1987)
\bibitem{Lyc1} V. V. Lychagin, Geometric singularities of solutions of nonlinear differential equations, Soviet Math. Dokl. 24 (3) 680-685  (1981)
\bibitem{Lyc2} V. V. Lychagin, Geometric theory of singularities of solutions of nonlinear differential equations, J. Soviet Math. 51(6) 2735-2757 (1990)
\bibitem{Mag} R. Magnanini, An introduction to the study of critical points of solutions of elliptic and parabolic equations,
Rend. Istit. Mat. Univ. Trieste 48 121-166  (2016)
\bibitem{Poi} H. Poincar\'e,  Sur les propri\'et\'es des fonctions d\'efinies par les \'equations aux diff\'erences
partielles: 1\`ere th\`ese Paris: Gautiers-Villars (1879)
\bibitem{Rak1} A. Kh. Rakhimov, Singularities of solutions of quasilinear equations, St. Petersburg Math. J. 4 (4), 217-224 (1992) 
\bibitem{Rak2} A. Kh. Rakhimov, Singularities of Riemann invariants, Funct. Anal. and Appl. 27, no. 1, 39-50  (1993) 
\bibitem{Rem1} A. O. Remizov, Implicit differential equations and vector fields with non-isolated singular points, Sb. Math. 193(11-12), 1671-1690 (2002)
\bibitem{Rem2} A. O. Remizov, A brief introduction to singularity theory, https://www.sissa.it/fa/download/publications/remizov.pdf 
\bibitem{Sul} B. I. Suleimanov, Cusp catastrophe in slowly varying equilibriums, JETP 95(5) pp. 944-956 (2002)
\bibitem{Sze} G. Szego, Orthogonal Polynomials, American Mathematical Society (1967)
\bibitem{Thom}  R. Thom,  Structural Stability and Morphogenesys, New York: Benjamin-Addison (1975)
\bibitem{W} H. Whitney, {On singularities of mappings of Euclidean spaces. I. Maps of the plane into the plane},
Ann. of Math. 62(3) 374-410 (1955)
\end{thebibliography}
\end{document}